\newcommand{\ThesisAuthorFirstName}{Steffen}
\newcommand{\ThesisAuthorFamilyName}{Schulz}
\newcommand{\ThesisAuthorImmatriculationNumber}{1166051}
\newcommand{\ThesisAuthorUniversity}{University of Siegen}
\newcommand{\ThesisAuthorFaculty}{Intelligent Systems Group}
\newcommand{\ThesisAuthorDegree}{Undergraduate}
\newcommand{\ThesisAuthorStudies}{Recommender Systems}
\newcommand{\ThesisAuthorStudiesSpecialization}{Algorithm Evaluation}
\newcommand{\ThesisAuthorRegulationVersion}{2012}

\newcommand{\ThesisTitle}{Algorithm Performance Spaces for Strategic Dataset Selection}
\newcommand{\ThesisType}{Bachelor's Thesis}
\newcommand{\ThesisStartDate}{19.06.2024}
\newcommand{\ThesisEndDate}{14.03.2025}

\newcommand{\ThesisExaminerATitle}{Prof.\,Dr.} 
\newcommand{\ThesisExaminerAFirstName}{Joeran}
\newcommand{\ThesisExaminerAFamilyName}{Beel}
\newcommand{\ThesisExaminerADepartment}{Intelligent Systems Group}

\newcommand{\ThesisExaminerBTitle}{M.\,Sc.} 
\newcommand{\ThesisExaminerBFirstName}{Lukas}
\newcommand{\ThesisExaminerBFamilyName}{Wegmeth}
\newcommand{\ThesisExaminerBDepartment}{Intelligent Systems Group}

\documentclass[12pt]{book}

\usepackage[T1]{fontenc}
\usepackage[utf8]{inputenc}

\usepackage[paper=a4paper, tmargin=3cm, bmargin=3cm, lmargin=2.5cm, rmargin=2.5cm, bindingoffset=1cm]{geometry}
\usepackage{tabularx}
\usepackage{graphicx}
\usepackage[super]{nth}

\usepackage{amsmath}
\usepackage{booktabs}  
\usepackage{url}  

\usepackage{hyperref}

\usepackage[acronym]{glossaries}
\makeglossaries
\newacronym{APS}{APS}{Algorithm Performance Space}
\newacronym{HPO}{HPO}{Hyperparameter Optimization}
\newacronym{nDCG}{nDCG}{Normalized Discounted Cumulative Gain}
\newacronym{PCA}{PCA}{Principal Component Analysis}
\newacronym{i.e.}{i.e.}{id est}
\newacronym{e.g.}{e.g.}{exempli gratia}

\usepackage{natbib}
\begin{document}

\pagestyle{empty}

{\centering
\par\vspace*{1cm}
{\scshape\large \ThesisAuthorFaculty{}\par\vspace{-0.15cm}}
{\scshape\LARGE \ThesisAuthorUniversity{}\par}
\vspace{1.0cm}
{\huge\bfseries \ThesisTitle{}\par}
\vspace{1.0cm}
{\scshape\LARGE \ThesisType{}\par}
{\itshape\large \ThesisAuthorStudies{}\par\vspace*{-0.07cm}}
{\itshape\large \ThesisAuthorStudiesSpecialization{}\par\vspace*{-0.07cm}}
{\itshape\LARGE \ThesisAuthorFirstName{} \ThesisAuthorFamilyName{}\par}
{\itshape\large \ThesisAuthorImmatriculationNumber{}\par}
\vspace{1.0cm}
{\ThesisEndDate{}}
\vfill

\includegraphics[width=0.5\textwidth]{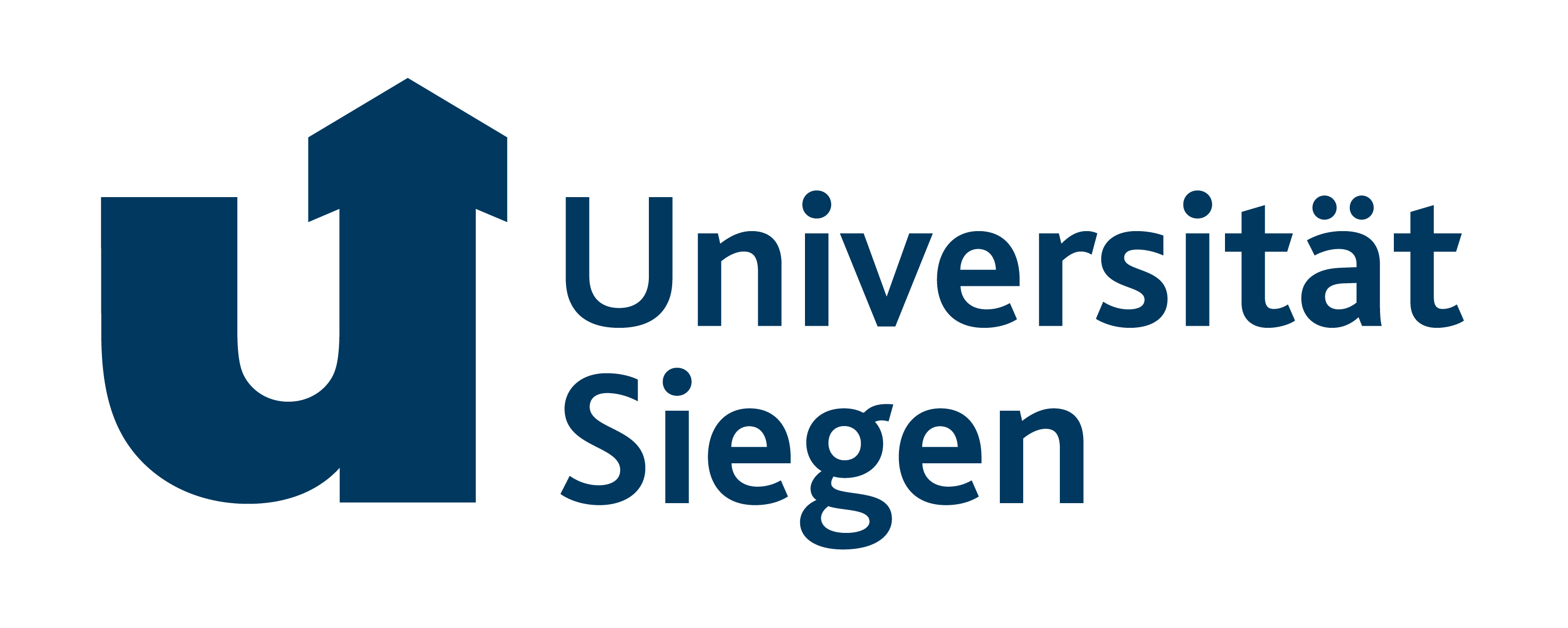}\par\vspace{0.1\textwidth}
\begin{figure}[h]
    \centering
    \includegraphics[width=0.25\textwidth]{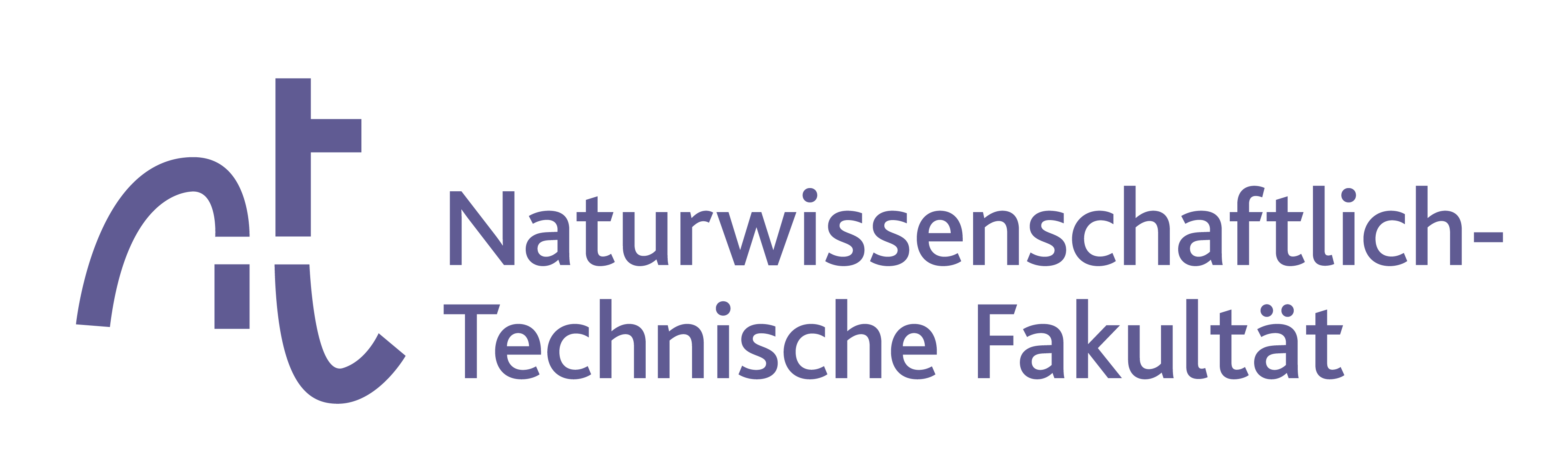}
    \includegraphics[width=0.1\textwidth]{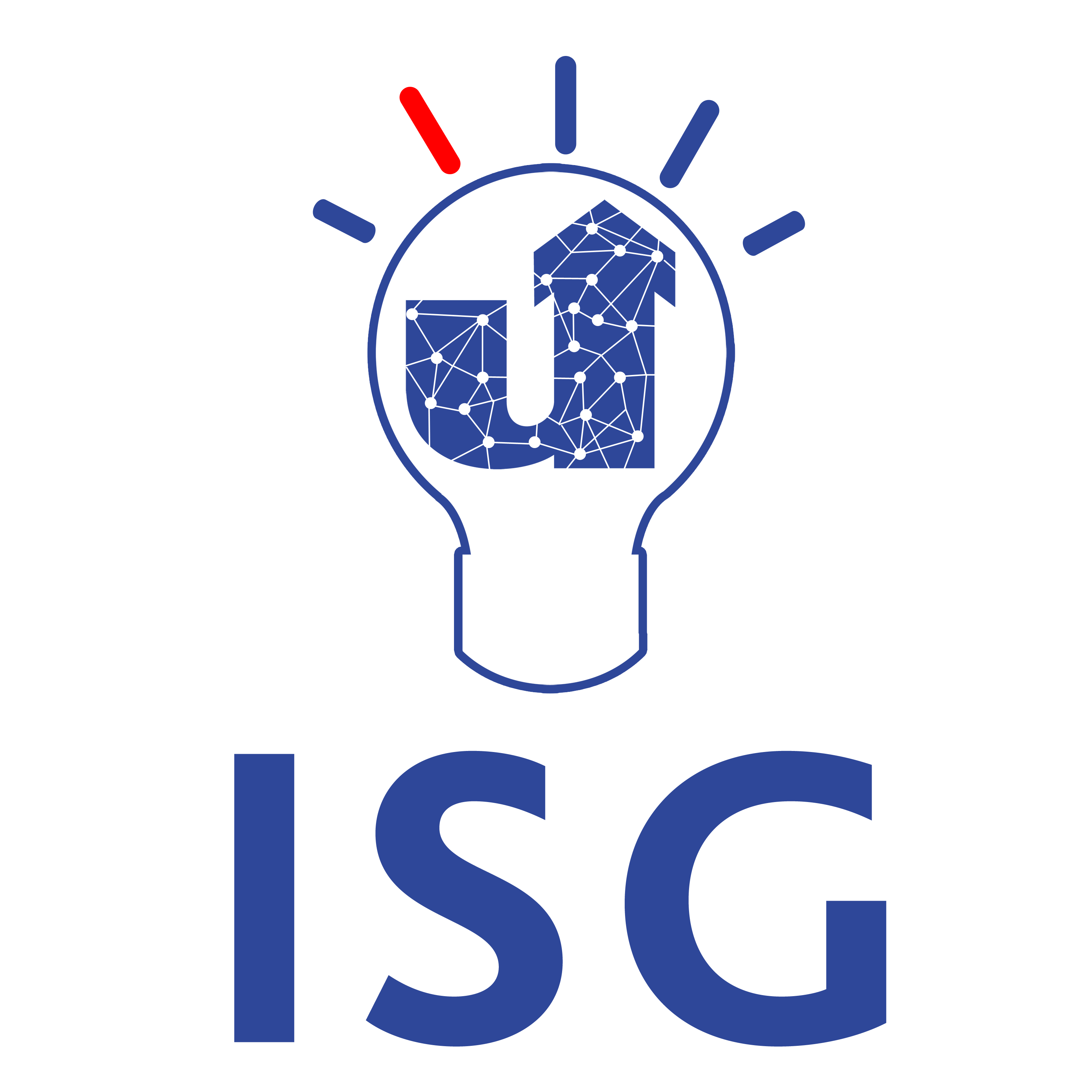}
\end{figure}

\begin{tabularx}{\textwidth}{@{} l X r @{}}
	{\small \textbf{Examiners}} & &  \\
	{\ThesisExaminerATitle{} \ThesisExaminerAFirstName{} \textsc{\ThesisExaminerAFamilyName{}}} & & {\ThesisExaminerADepartment} \\
	{\ThesisExaminerBTitle{} \ThesisExaminerBFirstName{} \textsc{\ThesisExaminerBFamilyName{}}} & & {\ThesisExaminerBDepartment} \\
\end{tabularx}\par}

\clearpage{}

\noindent This \ThesisType{} is handed in according to the requirements of the \ThesisAuthorUniversity{} for the study program \ThesisAuthorDegree{} \ThesisAuthorStudies{} of the year \ThesisAuthorRegulationVersion{} (PO\,\ThesisAuthorRegulationVersion{}).\par

\vfill

\begin{tabularx}{\textwidth}{@{} l X @{}}
	{\small \textbf{Process period}} & \\
	{\ThesisStartDate{} to \ThesisEndDate{}} & \\[0.4cm]
	{\small \textbf{Examiners}} & \\
	{\ThesisExaminerATitle{} \ThesisExaminerAFirstName{} \textsc{\ThesisExaminerAFamilyName{}}} & \\
	{\ThesisExaminerBTitle{} \ThesisExaminerBFirstName{} \textsc{\ThesisExaminerBFamilyName{}}} & \\
\end{tabularx}\par

\clearpage{}

\pagenumbering{Roman}
\pagestyle{plain}
\phantomsection
\addcontentsline{toc}{chapter}{Abstract}
\section*{Abstract}
The evaluation of new algorithms in recommender systems frequently depends on publicly available datasets, such as those from MovieLens or Amazon. Some of these datasets are being disproportionately utilized primarily due to their historical popularity as baselines rather than their suitability for specific research contexts. This thesis addresses this issue by introducing the Algorithm Performance Space, a novel framework designed to differentiate datasets based on the measured performance of algorithms applied to them. An experimental study proposes three metrics to quantify and justify dataset selection to evaluate new algorithms. These metrics also validate assumptions about datasets, such as the similarity between MovieLens datasets of varying sizes. By creating an Algorithm Performance Space and using the proposed metrics, differentiating datasets was made possible, and diverse dataset selections could be found. While the results demonstrate the framework's potential, further research proposals and implications are discussed to develop Algorithm Performance Spaces tailored to diverse use cases.

\renewcommand*\contentsname{Table of Contents}
\phantomsection
\tableofcontents
\addcontentsline{toc}{chapter}{Table of Contents}


\let\cleardoublepage\clearpage
\listoffigures
\addcontentsline{toc}{chapter}{List of Figures}
\listoftables
\addcontentsline{toc}{chapter}{List of Tables}
\printglossaries
\addcontentsline{toc}{chapter}{Acronyms}

\chapter{Introduction}
\pagenumbering{arabic}
\pagestyle{plain}
\setcounter{page}{1}
\section{Background}
In modern society, we are constantly exposed to products and services powered by machine learning algorithms, categorized under recommender systems. The enhancement of these recommender systems has become a focal point for both industry and research, with ongoing innovation and novel algorithms emerging routinely. Naturally, these novel algorithms require evaluation to demonstrate their effectiveness, performance, or other critical metrics relevant to their intended application. Given that most recommender systems are based on machine learning algorithms, including deep learning algorithms, analytical evaluation can be challenging. Therefore, researchers train, test, and evaluate their validity using large datasets, with the aim of predicting their performance on new and unseen data in the future.

In recent history, concerns about the evaluation process has been raised by researchers, as highlighted by \cite{ferrari2019we}. One of the primary issues involves the reproducibility of results, with 11 of 18 recent studies introducing algorithms not being reproducible. One of the highlighted reasons lies in the reliance on datasets that are either not publicly available or insufficiently documented. Moreover, the inconsistent pre-processing of datasets, coupled with arbitrary train-test splits, leads to skewed evaluations that can over exaggerate the performance of novel methods. \cite{bauer2024exploring} found in their literature review that only a handful of all available datasets were used for offline evaluation, especially the MovieLens and Amazon datasets. MovieLens datasets were used in 32 of the 57 reviewed articles published between 2017 and 2022. The Amazon datasets were used in 24 papers. Similar findings are presented and discussed by \cite{sun2020we, beel2024informed, chin2022datasets, beel2019data, sun2022daisyrec}.

\section{Research Problem}
Although it may not necessarily be a problem to evaluate recommender-systems algorithms on only a handful of well-known datasets, the justification for doing so is often weak or missing entirely. \cite{cremonesi2021progress} attest to this in their analysis, stating that not providing a rigorous justification for the selection of the datasets is deemed "acceptable" by the research community. The study conducted by \cite{beel2024informed} analyzed 41 full papers from the ACM RecSys 2023 conference that utilized offline evaluations. They found that none of the authors offered a direct and detailed justification for their use of datasets. 18 of the 41 papers explained their choices of datasets by labeling them either as "benchmark" or "widely used". Some (13 of 41) mentioned the domain of application or the type of recommendation, which would require specific datasets with additional information.

A careful and well reasoned offline evaluation setup can be crucial to properly justify progress by presenting new algorithms. \cite{rendle2019difficulty} demonstrated, that fine-tuning algorithms and pre-processing steps for specific datasets such as the MovieLens 10M dataset can lead to significant performance increases, even for algorithms commonly used in baselines like Bayesian MF or SVD++. Similarly \cite{zhao2022revisiting} conducted an extensive study on carefully fine-tuned algorithms from the RecBole library on eight datasets. They demonstrated significant differences in performance for different setups and datasets, highlighting the need to properly adjust baseline algorithms.

\section{Thesis Goal}
Accordingly, the dataset selection for evaluating new algorithms should be under similar scrutiny to ensure that the presented results are reproducible and meaningful. This thesis attempts to provide a new perspective for dataset selection by introducing Algorithm Performance Spaces as a tool to justify the choices made. By doing so, datasets and selections of datasets could then be described and differentiated in an additional way, based on their evaluated performance on algorithms featured in the Algorithm Performance Space.

\chapter{Background}
\section{Datasets}
To evaluate machine learning algorithms, we usually depend on training it with large amounts of data and measuring its performance on another subset of data, which was not used in training, emulating future unseen data. There are multiple ways for these algorithms to learn; in recommender systems, we are looking at a special case of supervised learning. In a supervised learning environment, the data to train and evaluate algorithms is essentially a list of details about decisions, representing all the inputs into the algorithms, with the expected outcome already known. In this way, the algorithm would process the input and compare the results with the expected output, measuring the difference between the result and expectation by calculating an error term. The error will then be minimized iteratively by doing this process for all individual cases in the training dataset. This can be repeated multiple times, each iteration being called an epoch. After training, the algorithm can then be evaluated by presenting it with new data that was not used in the training process. The results of this evaluation will be used to calculate various metrics to measure it's performance.

In recommender systems, individual cases in a dataset are usually called interactions, which feature at least the user and item involved. The interaction describes, depending on the context, that this user interacted with the item in some way. For example, this could be a customer in a web shop clicking on a specific product. Each interaction can come with numerous other information, such as the time it happened or how expensive the product was. By training on a dataset with a lot of these interactions, the algorithm will then ideally learn, what items a user would like to interact with in the future. Recommender Systems algorithms use diverse methods to achieve this, like Collaborative Filtering (\cite{sarwar2001item}), Matrix Factorization (\cite{koren2009matrix}) and many more. The makeup of an interaction can also be very diverse, depending on the context and origin of the data. Interactions with an explicit output, such as a rating on a rating scale, are called explicit feedback data. In contrast, interactions without any specified output are often referred to as implicit feedback data, since the desired outcome is implied to be positive for all interactions in a dataset. There are numerous other ways for recommender systems to work, such as with emphasis on time and sequences, which will not be subject of this thesis.

\section{Algorithm Training and Evaluation}
To find interactions to train a recommender algorithm, the domain in which the algorithm is supposed to work in is considered. This means for a company that runs a movie streaming service such as Netflix, that it could use it's own real-world data to iteratively improve their algorithms. In a broader context and in research, new and improved algorithms are constantly developed, that could potentially work in all kinds of domains, being capable of general recommendations. In research, the access to real-world data of active companies is limited, since the performance of a recommender system, and subsequently the data to improve them, is an asset with value attached. Most often in research, publicly available datasets are used instead to train and evaluate algorithms. These datasets can come from different domains and sources and are released to help researchers, for machine learning competitions or others reasons. There can be huge differences in how data was collected, what information is included and who released it.

Using as many datasets as possible to train algorithms would be desirable. With lots of data from diverse backgrounds, better performance and more generalization could be ensured. Unfortunately as of today, the computational resources available for research is limited. The process of training and evaluating algorithms can be quite costly in that regard, and doing so with potentially hundreds of dataset is simply out of question in almost all cases. Usually, only a handful of datasets are used instead, to acquiesce to the limitations of time and computational power.

Different datasets can vary a lot when it comes to additional information, different data formats, size, uniqueness and others. We need to pre-process them to make them compatible with an algorithm that expects it's inputs to be in the same format, and to be able to compare the results later on. Some methods are quite self-explanatory, such as removing interactions with missing data or removing duplicates. Some are done to improve the calculation speed of the algorithm by normalizing entries and converting full-text names into serialized numbers. There are other commonly used methods which are more controversial, since they can alter datasets significantly, calling their representation for real-world data into question (\cite{beel2019data}). In pruning, users and/or items are removed from the dataset, when they have less then a certain number of occurrences in the dataset. In addition, datasets can also be converted to meet certain requirements, such as converting explicit rating based interactions into implicit feedback interactions. 

When feeding an algorithm test data to evaluate it, performance metrics are calculated to rate the performance of the algorithm. In the context of recommender systems evaluation, this usually works in a ranking based system. Here, the algorithm produces a list of items it would recommend a given user, sorted by relevance. There are predictive metrics to evaluate the accuracy of the predicted items, or ranked-based metrics that factor in the order in which the items were sorted in the list. There are also several other metrics, such as diversity or coverage. The number of items in this list is variable and usually denoted as the parameter K.

\chapter{Related Work}
  

\citet{sun2020we} conducted a comprehensive review of 85 papers published across eight top-tier conferences in the field of recommender systems, such as RecSys, KDD, and SIGIR, from 2017 to 2019. Their analysis aimed to address critical challenges in ensuring reproducibility and fairness in the evaluation of recommendation algorithms. The review revealed two primary issues affecting dataset usage: domain diversity and version diversity. Domain diversity refers to the wide range of datasets used across different studies, often drawn from distinct domains like movies, music, and e-commerce, which complicates cross-study comparisons and generalizations. Version diversity highlights inconsistencies within datasets that share the same name but have been updated multiple times, such as the Yelp dataset, which has seen at least three significant iterations over time. The authors cataloged datasets, pre-processing strategies, and evaluation methodologies used in these studies. They observed that many papers lacked standardized protocols, with significant variability in dataset selection and reporting practices. In addition, only a subset of studies disclosed critical details regarding data filtering, splitting strategies, or parameter tuning, further compounding the reproducibility challenge. Their findings underscored the urgent need for systematic benchmarks, as inconsistencies in dataset handling often lead to contradictory or incomparable results.

\citet{fan2024our} conducted a detailed exploration of the MovieLens dataset to examine the relationship between its data generation mechanisms and the evaluation of recommendation algorithms. They highlighted significant discrepancies between the interaction contexts of MovieLens data and real-world recommendation settings. Specifically, they observed that nearly half of all users completed their ratings within a single day, often influenced by the platform’s internal recommendation algorithms. This temporal concentration and guided interaction process raise questions about the dataset's ability to generalize model performance to broader, more diverse recommendation scenarios. The study underscores the necessity of scrutinizing benchmark datasets to ensure their alignment with practical use cases, emphasizing that results obtained on MovieLens should not be over-relied upon when evaluating recommender systems for real-world applicability.

In their analysis, \citet{chin2022datasets} focus extensively on improving the dataset selection process for offline evaluation of recommender systems, aligning closely with the goal of this thesis. They highlight the arbitrary nature of dataset selection in much of the current literature. To show this, they first systematically analyzed 45 publicly available datasets used in 48 papers from five top-tier conferences, such as RecSys and SIGIR, from 2016 to 2020. Their investigation revealed substantial variability in how datasets are utilized, with only 24\% of datasets being used in five or more studies, while over half were used in just one paper.
To address this, \citet{chin2022datasets} then used five characteristics of the dataset: space, shape, density, and interaction distributions between users and items. By applying k-means clustering, they grouped 51 datasets into five clusters, selecting three representative datasets from each cluster for evaluation with five algorithms (UserKNN, ItemKNN, RP3beta, WMF, and Mult-VAE). Their empirical study demonstrated that dataset characteristics significantly impact algorithmic performance. For instance, RP3beta exhibited a 45\% performance improvement on sparse, moderate-sized datasets compared to dense, larger ones, where it underperformed relative to models like UserKNN and Mult-VAE.


Solely relying on data characteristics however may not be sufficient to predict algorithm performance on similar datasets. \citet{chin2022datasets} observed that datasets from different clusters occasionally showed comparable performance patterns, while those within the same cluster often produced inconclusive results. For example, the Amazon dataset (Movies \& TV) from Cluster 1 demonstrated performance closely aligned with the Amazon dataset (Toys \& Games) from Cluster 3. In both cases, RP3beta consistently outperformed other algorithms, while Mult-VAE ranked the lowest, indicating similar algorithmic trends despite being in different clusters. In Cluster 4, no clear pattern emerged, as the top-performing and lowest-performing algorithms varied across datasets. For instance, when looking at the ML-100k and the Amazon (Musical Instruments) datasets, the rankings of algorithms were reversed, highlighting stark differences in their relative effectiveness.

Research by \citet{beel2016towards}, \citet{ferrari2019we}, and \citet{cremonesi2021progress} also suggest that while dataset characteristics can be important, they alone are unable to fully determine an algorithm's effectiveness. For that reason, \citet{beel2024informed} introduced the idea of differentiating datasets based on algorithm performance alone, which is the pre-cursor and basis of this thesis.

\chapter{Methodology}
\section{Algorithm Performance Space}
The concept of the Algorithm Performance Space (\textbf{APS}) was initially introduced by \citet{beel2024informed} as an extension of the Algorithm Performance Personas established by \citet{tyrrell2020algorithm} to select algorithms per dataset instance for meta-learning. In this Algorithm Performance Space, datasets are represented as points located in it. The dimensions used to create the APS represent different recommendation algorithms and their respective measured performance when training and evaluating these datasets. The algorithms used, as well as the specific performance metric, are not invariable. With these points in the APS, we can now analyze and differentiate between them, and make informed decisions about dataset selections. The rational that makes conclusions from the APS meaningful, is the same reason machine learning is used in general. Instead of figuring out, which dataset characteristics are more indicative than others for the resulting performance, the APS directly consults the results of trained recommendation models, disregarding how the algorithms arrived there. This way, the APS should be able to create a meaningful distinction between a dataset, that all algorithms performed great on, from a dataset that yielded poor performance, regardless of what led to that result.

\begin{figure}
    \centering
    \includegraphics[width=1.0\linewidth]{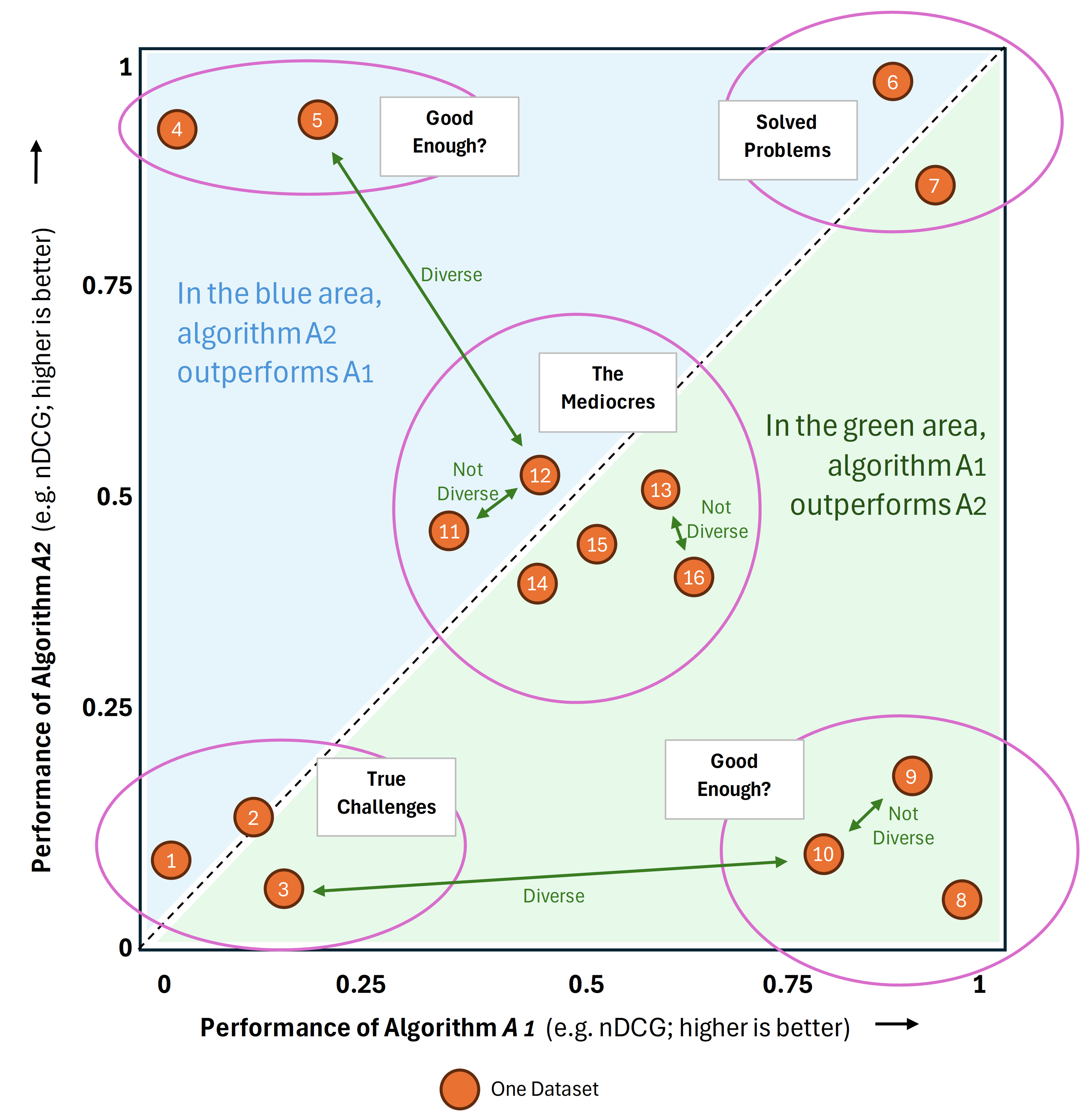}
    \caption{Illustration of the Algorithm Performance Space (APS) as introduced by \cite{beel2024informed}. The two axes represent the measured performance of algorithms \textit{A\textsubscript{1}} and \textit{A\textsubscript{2}}. The orange points 1 to 16 represent datasets. These datasets are clustered based on similar performance on algorithms \textit{A\textsubscript{1}} and \textit{A\textsubscript{2}} and can now be differentiated.}
    \label{fig:exampleAPS}
\end{figure}

In the example shown in Figure \ref{fig:exampleAPS}, the Algorithm Performance Space is shown as a two-dimensional space to explain the concept. The two dimensions represent algorithms A\textsubscript{1} and A\textsubscript{2} ranging from 0 to 1, which stand for the measured performance for a particular instance. Each of the 16 points in this space represents a certain dataset that the algorithms trained on. This means for example, that the dataset for point P\textsubscript{3}, located around the origin, got poor performance from both algorithms A\textsubscript{1} and A\textsubscript{2}, while for dataset P\textsubscript{10}, algorithm A\textsubscript{1} instead performed great on. So based on the measured performance differences, this indicates a high degree of diversity between these two datasets. Conversely, datasets P\textsubscript{10} and P\textsubscript{9} both feature similar measured performance, making them not very diverse. Datasets around the diagonal from the origin in the bottom left, towards the top right corner of the space all feature similar results for both algorithms A\textsubscript{1} and A\textsubscript{2}. The APS also enables interpretation of different areas and clusters. Datasets located in the top right corner for example got great performance from all algorithms and could therefore be considered as \textit{"solved problems"}. Meanwhile, datasets around the origin can be described as \textit{"true challenges"}, since no algorithm was able to sufficiently solve the recommendation problem for them.

The different areas of the APS cannot really be seen as equals. Datasets in the top right corner can be considered "easy", since all algorithms featured in the APS solved the recommendation problem for them. Therefore, presenting similar good performance on them with a new algorithm would not be of great significance. Conversely, when a new algorithm would be able to perform great on datasets in the bottom left corner, it would indicate great progress. In contrast to the two-dimensional APS in Figure \ref{fig:exampleAPS}, a proper APS would feature many algorithms and therefore be a high-dimensional space. This means, that besides the main diagonal, there can be way more than just the two clusters at the bottom right and top left corner of the illustration. There could be datasets clustered for algorithms A\textsubscript{1} - A\textsubscript{n-1} with a poor performance, and algorithm A\textsubscript{n} with a great performance. Alternatively, another cluster could exist for A\textsubscript{1} - A\textsubscript{n/2} with poor performance and A\textsubscript{n/2+1} - A\textsubscript{n} with great performance. Datasets from these "high-variance" clusters could also be of interest for targeted experiments, based on the algorithms involved.

When considering a potential new algorithm, the diversity of datasets based on the APS can be informative for its evaluation. Since for a cluster of non-diverse datasets, all algorithms performed consistently similar (not necessarily equal) on all of them, it can be expected that the new algorithm would show the same consistency on these datasets. Due to that consistency, it would therefore be sufficient to only include one or two of those non-diverse datasets, as the similar performance on the other datasets of the cluster can be inferred. When forced to select only a handful of datasets for evaluation, as time and compute are limited resources, a highly diverse dataset selection would be more informative, as it would be representing an even broader set of datasets not included in the selection. However, this does not always have to be the case, depending on the research goal. It could also be reasonable to only feature non-diverse datasets of a specific cluster in a selection. For example, a potential new algorithm could deliberately try to differentiate between these datasets.

As mentioned by \cite{beel2024informed}, these examples of use-cases and arguments are not meant to be definitive recommendations of specific datasets. The APS merely serves as a concept to make a more meaningful selection of dataset possible. This extends towards the details of metrics and the experiment in the following sections. The algorithms chosen, metrics measured, datasets used - these are all factors that create a unique APS targeted towards a specific use-case, leaving researchers with the responsibility to provide individual argumentation for a dataset selection based on the APS.

\section{Visualization with Mini-APS and PCA}
While the illustration in Figure \ref{fig:exampleAPS} serves as a great example to explain the concept, in reality the APS is a high-dimensional space. For the purposes of analyzing the intricacies of the APS, it is helpful to convert it in a two-dimensional format. Although it cannot capture the full picture of the entire APS, even a specific subset of it can provide valuable insights into the value it provides.

\textbf{Mini-APS}, instead of trying to visualize the entire APS, focus on two chosen dimensions of the high-dimensional space. This means that from all n algorithms featured in the APS, two are selected to display them on a two-dimensional graph, along with all the points in it, representing the performance on the datasets. The resulting graph essentially looks similar to the illustration in Figure \ref{fig:exampleAPS}. To visualize the whole APS in this way for \(n\) algorithms, \(n*(n-1)\) mini-APS have to be created, one for each algorithm pairing possible. In practice and in the context of this thesis, the resulting mini-APS will feature the performance results as normalized between 0 (as is) and 1 (for the highest measured performance) to improve visibility.

Since Mini-APS can only show a small subsection of the APS, to visualize the entire APS, the high-dimensional space can be reduced using Principal Component Analysis (\acrshort{PCA}). This is achieved by calculating the direction, where the points in the space have the highest variance, \acrshort{i.e.} the greatest spread, and assigning it as the first component. In case of a reduction into a two-dimensional space, the second component is then determined again, by finding the direction with the highest variance that is orthogonal to the first component. Each component explains a certain portion of the total variance in the original space. These two components can now be displayed on a graph and used to analyze the APS. It is important to understand that these components are deprived of the original meaning of algorithms with their performance metrics. Furthermore, the explained variance for the two components can be uneven, resulting in a wrong perception of the graph.

\section{Metrics}
Although Mini-APS and \acrshort{PCA} help visualize the APS, we want to distill the information in the APS into a more usable form. To allow researchers to explain their choice of datasets, it would be troublesome to rely on figures of graphs to do so. For this reason, three new metrics that are derived from the APS are introduced.
\begin{enumerate}
\item \textbf{Difficulty}
The \textit{Difficulty\textsubscript{APS}} metric is the most straightforward and obvious measure that can be derived from the APS. It can be calculated for each dataset individually and fundamentally reflects the degree to which recommendations for that particular dataset were correct. Visually speaking, it ranges from the origin in the bottom left corner to the top right corner of the illustration shown in Figure \ref{fig:exampleAPS}. In addition, a gradient depicting how a dataset would rank in this metric is shown in Figure \ref{fig:diffGradient}. It is calculated as follows:
\[Difficulty_{APS} = \frac{1}{n}\sum_{i = 1}^{n} x_i\]
Here, \(i\) represents one of the algorithms used to construct the APS, while \(n\) is the total number of algorithms used for the APS, \acrshort{i.e.} the dimension of the APS. \(x_i\) stands for the calculated performance metric for a given dataset-algorithm pairing.
\begin{figure}
    \centering
    \includegraphics[width=0.6\linewidth]{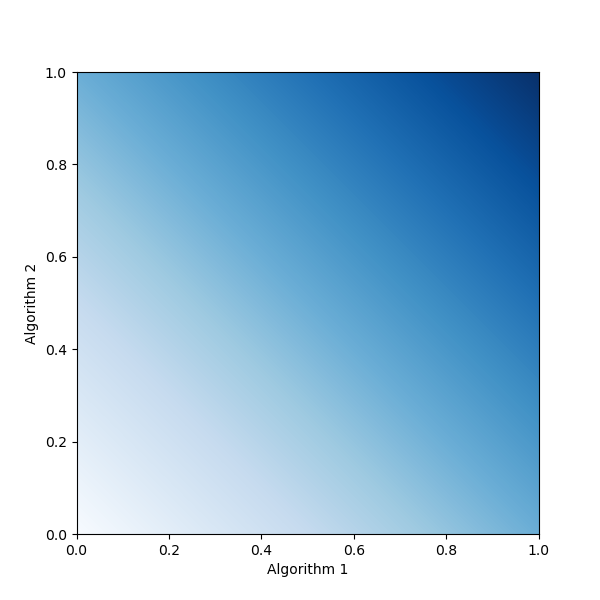}
    \caption{Visual representation of difficulty metric on a two-dimensional APS. Axes represent different algorithms from 0 to 1 in the chosen performance metric.}
    \label{fig:diffGradient}
\end{figure}
\item \textbf{Variance}
The \textit{Variance\textsubscript{APS}} metrics aims to describe, how consistent or inconsistent all algorithms performed on a given dataset. For example, a dataset D\textsubscript{1} on which the algorithms A\textsubscript{1} and A\textsubscript{2} of the APS performed well, algorithm A\textsubscript{3} performed mediocre, and algorithms A\textsubscript{4} and A\textsubscript{5} did under-perform, would be considered high in variance and would rank high in this metric. Conversely, dataset D\textsubscript{2} with similar performance for most if not all algorithms would rank low. In the visual representation of the APS in Figure \ref{fig:exampleAPS}, \textit{Variance\textsubscript{APS}} would be 0 on the diagonal spanning from the bottom left to the top right, while reaching 1 in the top left and bottom right corners respectively. Similar to the previous metric, a gradient describing the ranking visually in a two-dimensional space is shown in Figure \ref{fig:varGradient}. It is calculated as follows:
\[Variance_{APS} = \frac{2}{n(n-1)} \sum_{i=1}^{n} \sum_{j=i+1}^{n} |x_i - x_j|\]
The formula calculates the mean of the sum of all absolute distances between unique pairs of algorithms. Let $x_1, x_2, \dots, x_n$ represent the performance of each algorithm for the dataset. The double summation $\sum_{i=1}^{n} \sum_{j=i+1}^{n}$ ensures that all unique pairs of values $(x_i, x_j)$ are considered, where $i<j$ to avoid duplicate pairs. For each pair, the absolute difference $|x_i - x_j|$ is computed. The sum of these absolute differences is then divided by the total number of unique pairs, which is $\frac{n(n-1)}{2}$, to obtain the mean.
\begin{figure}
    \centering
    \includegraphics[width=0.6\linewidth]{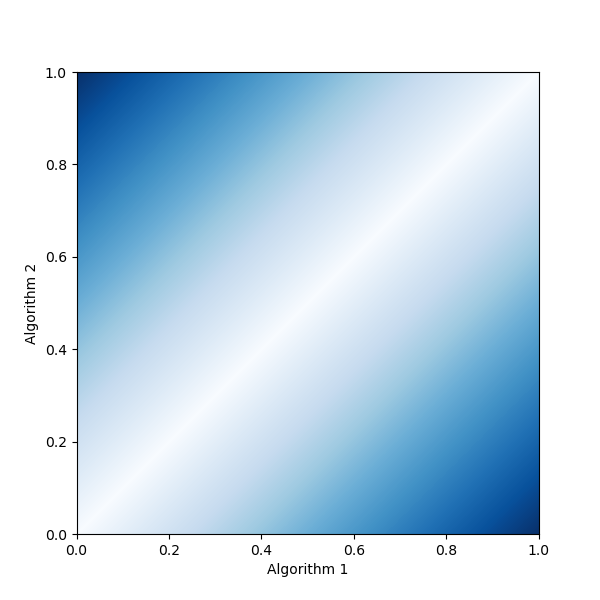}
    \caption{Visual representation of variance metric on a two-dimensional APS. Axes represent different algorithms from 0 to 1 in the chosen performance metric.}
    \label{fig:varGradient}
\end{figure}
\item \textbf{Diversity}
The previous two metrics describe datasets based on the different areas in the APS it could be located in. However, it is calculated and presented as a per-dataset metric. To be able to evaluate a selection of datasets, the relation between them needs to be described as well. The \textit{Diversity\textsubscript{APS}} metric explains this relationship by summarizing, how "different" each dataset performed in the APS, compared with other datasets in a given selection. For example, given a selection of three dataset D\textsubscript{1}, D\textsubscript{2} and D\textsubscript{3} and the APS consisting of algorithms A\textsubscript{1} and A\textsubscript{2}. Algorithm A\textsubscript{1} performed well on datasets D\textsubscript{1} and D\textsubscript{2} and poorly on D\textsubscript{3}, while algorithm A\textsubscript{2} performed well on datasets D\textsubscript{2} and D\textsubscript{3}, poorly on D\textsubscript{1}. In a visual representation, all three datasets would be located in different "corners" of the two-dimensional APS. The Diversity of this dataset selection, in terms of algorithm performance, can be considered high. To calculate this metric, we combine two aspects: the pairwise distance variance, which reflects the distribution of points relative to one another, and the coverage of the bounding box enclosing the points, which ensures the metric accounts for the spatial extent of the distribution. The mathematical formulation is given as:
\[
Diversity_{APS} = \left( 1 - \frac{\mathrm{Var}(\mathbf{D})}{\mathrm{MaxVar}} \right) \cdot \sqrt{\prod_{i=1}^{n} (x_{i,\text{max}} - x_{i,\text{min}})}
\]

\textbf{Where:}

\begin{align*}
\mathrm{Var}(\mathbf{D}) & = \frac{1}{N} \sum_{i=1}^N (x_i - \mu)^2, \\
\mathrm{MaxVar} & = \frac{(d_{\text{max}} - d_{\text{min}})^2}{4}, \\
d_{\text{max}} & = \sqrt{n}, \\
d_{\text{min}} & = 0. \\
\end{align*}

Given \(\mathbf{D}\) is the set of pairwise Euclidean distances between the points, \(\mathrm{Var}(\mathbf{D})\) is the variance of these distances, and \(\mathrm{MaxVar}\) represents the theoretical maximum variance for pairwise distances in the given space. The term \(1 - \frac{\mathrm{Var}(\mathbf{D})}{\mathrm{MaxVar}}\) normalizes and inverts the pairwise distance variance into the interval \([0, 1]\), such that lower variance translates to a more evenly distributed point set, while higher variance indicates clustering. The bounding box is defined by the minimum and maximum values of the points along each dimension, with its volume computed as \(\prod_{i=1}^d (x_{i,\text{max}} - x_{i,\text{min}})\). To reduce the disproportionate influence of very small volumes and to smooth the coverage contribution, the square root of this volume is taken. Multiplying the normalized variance by the transformed bounding box volume integrates these two components, resulting in a metric that penalizes clustering while rewarding both even spacing and maximal coverage.
\end{enumerate}

To demonstrate the effectiveness of the APS and the metrics introduced, we first conduct an experiment to construct the APS. Subsequently, we analyze the results through the lens of these three metrics. There are a few things to consider regarding these metrics. First, all of them are derived with the intention of using \acrshort{nDCG}, a normalized metric, as the performance metric of choice for algorithms in the \acrshort{APS}. Other metrics, especially when not normalized between 0 and 1, were not taken into account. Additionally, they capture only some of the features of the \acrshort{APS} that were of high interest, such as dataset diversity. Other researchers could value these features differently, and this freedom is intended, as described in 5.1. Ultimately, these metrics serve merely as a tool to put the results of the experiment in this thesis into perspective and gain insights into the validity of the concept of Algorithms Performance Spaces itself.

\section{Experiment}
To create and study the effectiveness of the Algorithm Performance Space (\acrshort{APS}), an experiment was conducted by training and evaluating 75 datasets on 5 recommendation algorithms. It is focused on the general recommendation problem with implicit feedback interactions. This means the processed datasets contain only a list of user-item interactions without added information, explicit ratings, or sequential information.

The datasets used in this experiment are listed in detail in Table \ref{table:resTable}. They are exclusively publicly available datasets, many of them often used in recent research history as shown by \citet{chin2022datasets}. 16 of the datasets are already based on implicit feedback, while the other 59 feature explicit ratings with various different rating scales. All 59 datasets with explicit ratings were converted into implicit ones, by treating all interactions with ratings over a certain threshold as implicit ones, while disregarding the rest. For example, the MovieLens datasets feature a rating scale of 1-5 stars to rate a Movie. All interactions with a rating of 4 and 5 were treated as a positive interaction and made implicit by removing the rating, leaving only user and item identifiers. All interactions with a rating of 1, 2 or 3 were deemed negative and therefore removed from the dataset. This process was done for all explicit datasets, even though the rating scales may differ. The practice of converting datasets in such a way is common practice and is well documented (\cite{hu2008collaborative, pan2008one}). The specific threshold used to categorize the explicit interactions into positive and negative ones is not uniformly agreed upon, in this experiment the threshold was set to 0.6.
Duplicate entries were removed, since giving the same user-item interaction more weight is not purposeful for a recommender system. Interactions containing invalid or missing information were also removed. As mentioned previously, all additional information, such as meta data about users or items, sequential information such as timestamps, or other meta data about the interaction was omitted, leaving only user and item identifier for each interaction.
Finally, all datasets were 5-core pruned. Pruning is a technique to remove outliers and noisy interactions from a dataset. By pruning a dataset we remove all interactions, where the user did not appear in at least 5 of them. The same removal process is done for items, that did not appear in at least 5 interactions. This 5-core pruning can significantly alter and reduce the size of a dataset, and there is discussion about its significance (\cite{beel2019data}). For the purpose of introducing the \acrshort{APS}, however, pruning, as well as many of the other implementation details were chosen to provide an example using well established practices.
The datasets were split into train- and test-sets with a holdout split of 80/20. The simple holdout split was favored over the more rigorous cross validation split because of the limitations of computational resources.

The following recommendation algorithms were used on all 75 datasets:
\begin{enumerate}
\item \textbf{BPR} (\cite{rendle2012bpr}): The Bayesian Personalized Ranking (BPR) algorithm optimizes personalized rankings by maximizing the posterior probability that a user prefers observed items over unobserved ones. It uses a pairwise ranking approach with stochastic gradient descent to learn latent user and item factors efficiently.

Hyperparameters:
\begin{itemize}
    \item \textbf{learning rate}: 5e-5, 1e-4, 5e-4, 7e-4, 1e-3, 5e-3, 7e-3
    \item \textbf{embedding size}: 32, 64, 128
\end{itemize}
\item \textbf{ItemKNN} (\cite{aiolli2013efficient}): Item-based k-Nearest Neighbors (ItemKNN) recommends items by identifying the most similar items to those a user has interacted with, based on a similarity metric like cosine similarity. It predicts user preferences by aggregating ratings or interactions from these similar items.

Hyperparameters:
\begin{itemize}
    \item \textbf{k}: 10, 20, 50, 100, 200
    \item \textbf{shrink}: 0.0, 0.1, 0.5, 1.0, 2.0
\end{itemize}
\item \textbf{MultiVAE} (\cite{liang2018variational}): Multinomial Variational Autoencoder (MultiVAE) employs a variational autoencoder framework, where an encoder maps user-item interactions into a latent distribution, and a decoder reconstructs the interactions from this latent space. It optimizes the model by minimizing the reconstruction error while regularizing the latent space to follow a prior distribution, typically Gaussian.

Hyperparameters:
\begin{itemize}
    \item \textbf{learning rate}: 5e-5, 1e-4, 5e-4, 7e-4, 1e-3, 5e-3, 7e-3
    \item \textbf{drop ratio}: 0.1, 0.2, 0.4, 0.5
\end{itemize}
\item \textbf{SGL} (\cite{wu2021self}): Self-supervised Graph Learning (SGL) enhances recommendation systems by augmenting the user-item interaction graph and learning node representations through contrastive learning. It maximizes the agreement between representations of the same node from different graph views while minimizing agreement between different nodes.

Hyperparameters:
\begin{itemize}
    \item \textbf{ssl tau}: 0.1, 0.2, 0.5
    \item \textbf{ssl weight}: 0.05, 0.1, 0.5
    \item \textbf{drop ratio}: 0.1, 0.2, 0.4, 0.5
\end{itemize}
\item \textbf{NeuMF} (\cite{he2017neural}): Neural Matrix Factorization (NeuMF) combines traditional matrix factorization with neural networks to model user-item interactions. It uses separate embeddings for users and items, which are fed into a neural network to capture non-linear relationships, enhancing the prediction of user preferences.

Hyperparameters:
\begin{itemize}
    \item \textbf{learning rate}: 5e-7, 1e-6, 5e-6, 1e-5, 1e-4, 1e-3
    \item \textbf{mlp hidden size}: [128, 64], [128, 64, 32], [64, 32, 16]
    \item \textbf{dropout prob}: 0.0, 0.25, 0.5
\end{itemize}
\end{enumerate}

All algorithms can be fine tuned by using various parameters for them. These parameters used to setup machine learning algorithms are called hyperparameters. By changing these hyperparameters, the resulting recommendation quality can vary drastically. There is no perfect set of hyperparameters for all datasets and use cases. Instead, we can attempt to find hyperparameters with various searching techniques for each use case. This process is called hyperparameter optimization (\acrshort{HPO}). Since the range of possible hyperparameters is theoretically infinite, we limit the space by using a set of predefined hyperparameters and chose among them in the optimization process. The hyperparameter space for each algorithm was largely influenced by the recommendations of the RecBole library by \cite{zhao2021recbole} and occasionally extended.

The recommendation algorithms were selected for several reasons. For one, they cover a great variety of different implementation methods, such as neighborhood-based, matrix-factorization, graph- or deep-learning-based techniques. They are also cheap in terms of computational requirements. Lastly, the space of hyperparameters to select from is relatively small, further reducing the need for an extensively long training process. 

To evaluate the trained models, the top-k recommendation list is generated for k=10. This list represents the k items that are recommended to a given user by the previously trained algorithm. The k items are the most likely options for the user to enjoy, according to the algorithm. The list is ordered by likelihood. With this list, the normalized discounted cumulative gain @ k (\acrshort{nDCG}@k) is calculated. The \acrshort{nDCG} measures the quality of the recommendation list by comparing it to the ground truth provided by the test-set, including its position in the list. This metric is normalized, meaning it ranges from 0 (ground truth item is not in the recommendation list) to 1 (ground truth item is at the very top of the list). According to \cite{gunawardana2012evaluating, cremonesi2010performance} the \acrshort{nDCG} metric is considered a useful metric to accurately describe the quality of a recommendation algorithm in a general setting.

The implementation of the specific training processes, as well as the evaluation, was handled by the RecBole library by \cite{zhao2021recbole}. This library was created by researchers to improve comparability and ease of use. The code used to run the training and evaluation can be found in Appendix 1.

The training and evaluation took place on the GPU-cluster of the University of Siegen. For each combination of algorithm and dataset, 20 hyperparameter combinations were run in 50 epochs for a maximum of 12 hours. The search algorithm for finding hyperparameter combinations was Tree-structured Parzen Estimators (TPE) by \cite{bergstra2011algorithms}, which models the probability distribution of good and bad combinations and iteratively sample the more promising regions. If the time to train and evaluate an algorithm on a dataset exceeded the limit of 12 hours, the process was stopped. If in that time frame, at least one full iteration with a hyperparameter combination was finished successfully, the iteration with the highest resulting \acrshort{nDCG} was chosen as the final result.

The \acrshort{nDCG}@10 for all algorithm-dataset pairings, were included in the \acrshort{APS}. To visualize the resulting space, we split the \acrshort{APS} into several Mini-\acrshort{APS}, each featuring only two algorithms. To allow for a more comprehensive visual overview of the \acrshort{APS}, we performed a principal component analysis (\acrshort{PCA}) on it to reduce the dimensions from X to 2.

\chapter{Results}

 In the experiment, out of the possible \(75*5=375\) possible dataset-algorithm pairings, 268 (71.5\%) of them finished at least one full \acrshort{HPO} round. 168 (44.8\%) of these pairings did complete all scheduled 20 \acrshort{HPO} rounds. 107 (28.5\%) pairings did not finish once. All pairings with at least one finished \acrshort{HPO} round were included in the \acrshort{APS}. This includes 71 out of the initial 75 datasets, with the other four datasets (Amazon-Books, Amazon-Fashion, DoubanShort, MillionSong) not finishing with any of the algorithms.
 The best performing dataset-algorithm pairing was the Jester dataset on the MultiVAE algorithm with an \acrshort{nDCG}@10 of 0.502. All results can be found in the table in Appendix 2, including the calculated metrics \textit{Difficulty\textsubscript{APS}} and \textit{Variance\textsubscript{APS}} as described in the Methodology. The dataset with the highest \textit{Difficulty\textsubscript{APS}} score was the Amazon-Electronics dataset with 0.989, the lowest being the Jester dataset with 0.516. The mean \textit{Difficulty\textsubscript{APS}} over all 71 datasets is 0.886, the median is 0.921. The dataset with the highest \textit{Variance\textsubscript{APS}} score was the Epinions dataset with 0.303, while the Amazon-CDs-and-Vinyl dataset with 0.003 scored lowest. The mean \textit{Variance\textsubscript{APS}} for all 71 datasets is 0.03, and the median is 0.02.

 \section{Mini-APS}

 \begin{figure}
    \centering
    \includegraphics[width=0.8\linewidth]{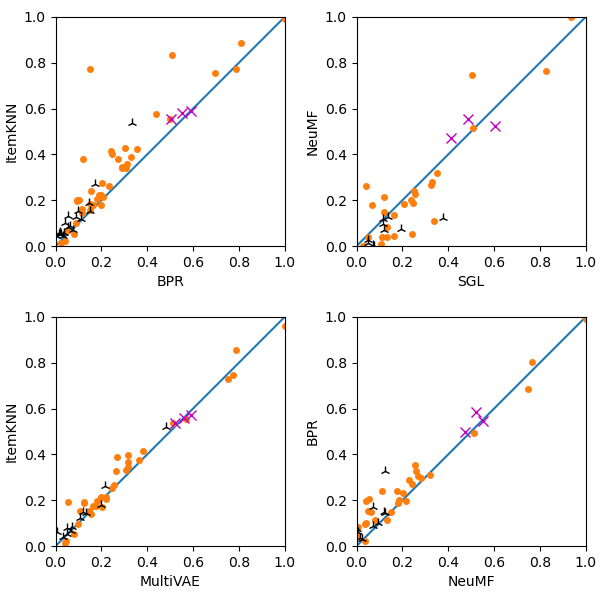}
    \caption{Four mini-APS as representation for the full \acrshort{APS}. MovieLens datasets are highlighted by violet markers, Amazon datasets by black ones. Axes denote normalized \acrshort{nDCG}@10 from 0 to 1 (best-performing \acrshort{nDCG} of $\approx$ 0.5).}
    \label{fig:miniAPSnew}
\end{figure}

 To get a better understanding of how the \acrshort{APS} is made up, a selection of 4 Mini-APS is shown in Figure \ref{fig:miniAPSnew}. All 25 Mini-APS can be found in Appendix 1. All Mini-APS are normalized between 0 for an \acrshort{nDCG} of 0, and 1 for the best measured nCDG of included results. In this case, the best measured \acrshort{nDCG} was roughly around 0.5 for all algorithms. Similarly to \cite{beel2024informed}, the MovieLens datasets are highlighted in pink, while the Amazon datasets are highlighted in black. In the top-left graph of Figure \ref{fig:miniAPSnew} the results for the algorithms ItemKNN and BPR are shown in a Mini-APS. Most of the datasets are near the diagonal and bottom-left corner, with a trend towards ItemKNN scoring higher and outliers stronger on ItemKNN than BPR. The MovieLens datasets are very close together in the center of the graph, while the Amazon datasets, while mainly clustered in the bottom-left, have outliers towards the center. In the top-right corner with the Mini-APS for the algorithms SGL and NeuMF, there is only a slight trend towards SGL in the bottom-left corner. Here, the MovieLens datasets are more apart, although still in the center of the graph. The Amazon datasets are clustered in the bottom-left corner, with a trend towards better performance on SGL, but without outliers in the center. The Mini-APS of MultiVAE and ItemKNN in the bottom-left show a strong alignment to the diagonal without any outliers towards one of the algorithms. The MovieLens cluster in the center is very small, similar to the first graph. The Amazon datasets, while aligned to the diagonal have a clear outlier in the center, right below the MovieLens datasets. Lastly the Mini-APS of NeuMF and BPR shows similar alignment to the diagonal with a slight trend towards the BPR algorithm and a few outliers in the top-right corner. MovieLens datasets are again clustered in the center, while the Amazon datasets without any strong outliers are located in the bottom-left corner with a trend towards BPR.

 \begin{figure}
    \centering
    \includegraphics[width=0.8\linewidth]{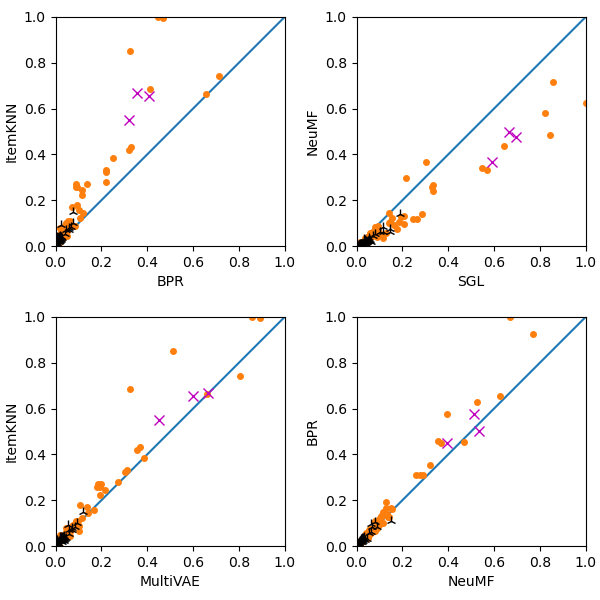}
    \caption{The same four mini-APS chosen in Figure \ref{fig:miniAPSnew} from \cite{beel2024informed} for comparison. MovieLens datasets are highlighted by violet markers, Amazon datasets by black ones. Axes denote normalized \acrshort{nDCG}@10 from 0 to 1 (best-performing \acrshort{nDCG} of $\approx$ 0.5).}
    \label{fig:miniAPSold}
\end{figure}

 To compare the \acrshort{APS} with the results of \cite{beel2024informed}, the Mini-APS featuring the same algorithms are shown in Figure \ref{fig:miniAPSold}. Compared to the \acrshort{APS} in Figure \ref{fig:miniAPSnew}, the cluster in the bottom-left corner of the \acrshort{APS} are more dense and focused between 0 and 0.2, while the new APS showing a broader cluster up to around 0.4 for most algorithms. The trend of ItemKNN outperforming BPR can be seen in both \acrshort{APS}, while the trend of SGL outperforming NeuMF can not be seen in the new \acrshort{APS}. In general, the cluster of the MovieLens datasets is slightly more spread out in the \acrshort{APS} of \cite{beel2024informed}. The Amazon datasets on the other hand are very close to each other without any strong outliers, in contrast to them being a bit more loosely connected and featuring outliers in the new \acrshort{APS}.

 \section{PCA Visualization}
 
 \begin{figure}
    \centering
    \includegraphics[width=1.0\linewidth]{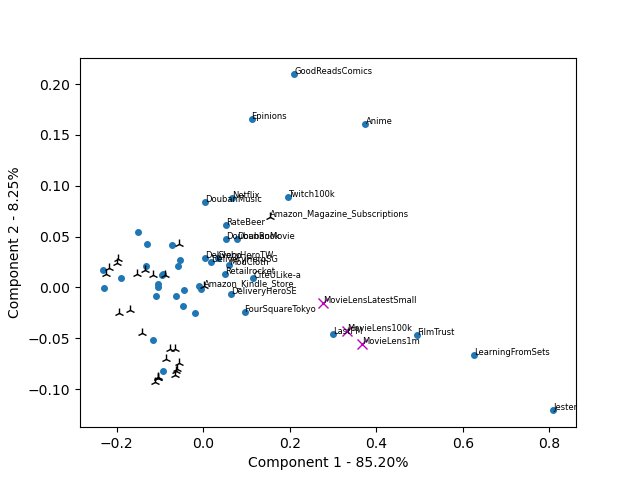}
    \caption{Results of dimensional reduction via \acrshort{PCA} to two dimensions. Axes show both calculated components with explained variance as percentage. MovieLens datasets are highlighted by violet markers, Amazon datasets by black ones.}
    \label{fig:PCAcompare}
\end{figure}

 To get a more holistic view of the \acrshort{APS}, the results of a dimensional reduction from 5 to 2 was performed with \acrshort{PCA}. The resulting diagram is shown in Figure \ref{fig:PCAcompare}. It is important to note, that the diagram in Figure \ref{fig:PCAcompare} is not shown in scale. The first dimension in the horizontal axis explains 85.2\% of the variance, while the second dimension in the vertical axis explains 8.25\%. Similarly to the Mini-APS, Amazon datasets are highlighted by black markers, MovieLens datasets by pink markers. The Amazon datasets are loosely clustered in the -0.2 to 0.0 range of component 1, with the exception of the Amazon Magazine Subscriptions dataset being close to 0.2. The MovieLens datasets are more closely grouped around 0.3 to 0.4 in component 1 and 0.00 to 0.05 in component 2. There are a few outliers for both axes. Namely, for component 1 the Jester, LearningFromSets and FilmTrust datasets can be located on the right with high values up to around 0.8. While the scale and variance for component 2 is much smaller, outliers up to 0.2 are located in the top of the graph, namely the GoodReadsComics, Epinions and Anime datasets. 
 
 \begin{figure}
    \centering
    \includegraphics[width=0.7\linewidth]{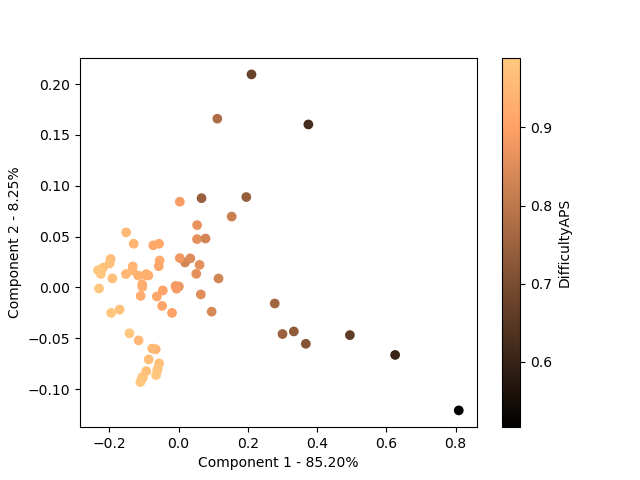}
    \caption{The results from the \acrshort{PCA} shown in Figure \ref{fig:PCAcompare} on the left, color-coded with \textit{Difficulty\textsubscript{APS}} calculated for each dataset.}
    \label{fig:PCAdiff}
\end{figure}
\begin{figure}
    \centering
    \includegraphics[width=0.7\linewidth]{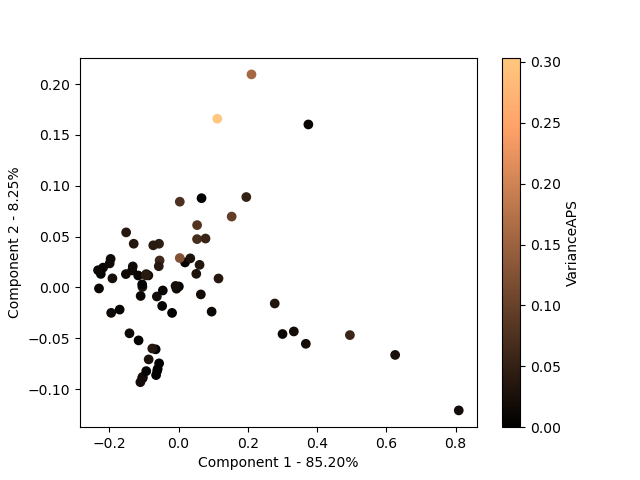}
    \caption{The results from the \acrshort{PCA} shown in Figure \ref{fig:PCAcompare} on the left, color-coded with \textit{Variance\textsubscript{APS}} calculated for each dataset.}
    \label{fig:PCAvar}
\end{figure}

 
 For the purpose of visualizing the calculated new metrics \textit{Difficulty\textsubscript{APS}} and \textit{Variance\textsubscript{APS}} are color-coded into the \acrshort{PCA}. In Figure \ref{fig:PCAdiff} this is done for the \textit{Difficulty\textsubscript{APS}} metric. Here, a clear correlation can be seen between component 1 in the x-axis and the \textit{Difficulty\textsubscript{APS}} metric. The calculated correlation coefficient between these two is \(\rho = 0.95\). In Figure \ref{fig:PCAvar} the same is done for the \textit{Variance\textsubscript{APS}} metric. Here, no strong correlation can be found. However, 2 of the strongest outliers in the \textit{Variance\textsubscript{APS}} metric can be found in the top of the graph with a higher value in component 2, namely the Epinions and GoodReadsComics datasets.

\section{Diversity of dataset selections}
To get an idea about how the results of the \acrshort{APS} would translate into the \textit{Diversity\textsubscript{APS}} metric, a few example sets of datasets were created. In Table \ref{table:resTable} a total of nine sets are shown with their respective calculated \textit{Diversity\textsubscript{APS}}. The Sets 0, 1 and 2 are the sets containing 2, 3 and 4 datasets respectively, and scored the highest in this metric with a \textit{Diversity\textsubscript{APS}} of 0.4698, 0.4468 and 0.4459. Similarly, the sets 3 to 5 are the least diverse sets, scoring 0.0002, 0.0059 and 0.0462 in \textit{Diversity\textsubscript{APS}}. Since the MovieLens and Amazon datasets were highlighted in the \acrshort{APS}, sets containing them are seen in the table sets 6 to 8. The MovieLens set containing the MovieLens1m, MovieLens100k and MovieLensLatestSmall datasets scored with 0.0399 very low on \textit{Diversity\textsubscript{APS}}. Similarly, set 7 containing the Amazon Arts \& Crafting, Digital Music and Gift Cards datasets is not very diverse with \textit{Diversity\textsubscript{APS}} of 0.0473. To demonstrate the impact of including outliers, like the best performing Jester dataset in a set, in the last set of the table is a supplemented version of set 7, including the Jester dataset. It scored relatively high on \textit{Diversity\textsubscript{APS}} with 0.3825. The best and worst combinations of sets were found by calculating \textit{Diversity\textsubscript{APS}} for all possible sets and picking the best/worst sets accordingly.

\begin{table}
\centering
\resizebox{\textwidth}{!}{
\begin{tabular}{llr}
\toprule
 & Dataset Selection & \textit{Diversity\textsubscript{APS}} \\
\midrule
0 & Jester, Food & 0.4698 \\
1 & Jester, Food, MovieLensLatestSmall & 0.4468 \\
2 & Jester, Food, Amazon-Magazine Subscriptions, FilmTrust & 0.4459 \\
3 & FourSquareNYC, MarketBiasModcloth & 0.0002 \\
4 & Amazon-Musical Instruments, Amazon-Prime Pantry, RentTheRunway & 0.0059 \\
5 & Amazon-Arts Crafts and Sewing, Amazon-Digital Music, Food, RentTheRunway & 0.0462 \\
6 & MovieLens1m, MovieLens100k, MovieLensLatestSmall & 0.0399 \\
7 & Amazon-Arts Crafts and Sewing, Amazon-Digital Music, Amazon-Gift Cards & 0.0473 \\
8 & Jester, Amazon-Arts Crafts and Sewing, Amazon-Digital Music, Amazon-Gift Cards & 0.3825 \\
\bottomrule
\end{tabular}
}
\caption{Table containing dataset selections with their respective score in \textit{Diversity\textsubscript{APS}}.}
\label{table:resTable}
\end{table}
 
\chapter{Discussion}
Not all pairings of datasets and algorithms finished all 20 scheduled \acrshort{HPO} runs, or did not finish at all. This leads to skewed results for certain datasets and algorithms. Especially large datasets, like the Amazon Books dataset or the Epinions dataset were not given enough time to train properly and find good hyperparameters for. In addition to that, some of the more complex and time-consuming algorithms, like NeuMF had less completed \acrshort{HPO} runs, which could lead to underestimating their performance in the \acrshort{APS}. This underscores how important the extent of time and resource investments must be in order to create a complete \acrshort{APS}, covering more algorithms and datasets and performing decent \acrshort{HPO} for every single one. The ranges of hyperparameters available to optimization was picked from the RecBole recommendations, these might not be optimal for every algorithm and, especially, every dataset. 32 datasets only feature results for a few, but not all algorithms. This further influences the validity of the \acrshort{APS}, and the derived metrics like \textit{Variance\textsubscript{APS}}. Looking at the Epinions dataset, it scored highest in \textit{Variance\textsubscript{APS}}, but only finished training for two of the five algorithms, indicating that the high discrepancy between the two trained algorithms could stem from unfinished \acrshort{HPO}. To avoid this, only datasets with results for all five algorithms were included in dataset selections in Table \ref{table:resTable}. The question however remains, if certain datasets would yield higher performance with better training and individual care for hyperparameters. Within the context of researchers showing under-appreciation of staple algorithms (\cite{zhao2022revisiting}), the claim the \acrshort{APS} wants to make, proofs to be difficult and requires in-depth considerations and care for each included algorithm and dataset. This reliance on quality results gets magnified even further, when taking into account the creation of multiple \acrshort{APS} for different kinds of recommendation problems, like \acrshort{e.g.} sequential or context-based recommendation.

That aside, even though not every single dataset-algorithm pairing yielded proper results, enough data for the \acrshort{APS} was generated nevertheless. The MovieLens dataset proved to perform very similar for all algorithms, indicating that relying solely on them for evaluating algorithms would not be meaningful for more than one of the datasets. The Amazon datasets did cluster to an extend in the lower \acrshort{nDCG}-ranges between 0.0 and 0.2, but feature a few outliers for certain algorithms. This indicates these datasets not being very diverse, similar to the MovieLens set. The outliers however go against this evaluation, showing that there could be meaningful differences for some of the Amazon datasets. Including these outliers, like the Amazon Magazine Subscription dataset from the \acrshort{APS} of this thesis, in a dataset selection only consisting of Amazon datasets would probably prove valuable. Conversely, specific dataset selections only including these outliers could skew the perception of performance on the majority of Amazon data. When looking at the Mini-APS in Figure \ref{fig:miniAPSnew} and Figure \ref{fig:miniAPSold} it is important to keep in mind that these are normalized graphs. Since all of the new Mini-APS are featuring the Jester dataset in the upper-right corner at 1.0, it's \acrshort{nDCG} performance only reached around 0.5 for all algorithm. This puts all shown Mini-APS essentially in the lower-left quarter of the actual \acrshort{APS}. However the expectations for an \acrshort{APS} on highly optimized training results would be, that the general makeup of spread and direction in these graphs would remain similar.

When comparing the Mini-APS to the results of \cite{beel2024informed}, they in general show a similar picture. The greater spread in the bottom-left corner of the \acrshort{APS} indicate more successful training runs. Especially the big cluster towards the origin in the previous results is likely due to unfinished or unsuccessful training, which the new \acrshort{APS} could approve upon. This could also mean, with even more sophisticated training, that the cluster would dissipate even more and spread, and wander towards the center. But even with the current results, picking truly challenging or unsolved datasets was made possible contrary to the results of \cite{beel2024informed}. The greater spread differentiates datasets that could yield improved performance from others that remained generally unsolvable. Additionally, the new results show stronger alignment to the diagonal from the bottom-left to the top-right. In the previous results, certain Mini-APS demonstrated stronger tendencies towards either one of the algorithms. This could indicate that these specific algorithms did not perform well, perhaps due to insufficient hyperparameter tuning or limited training time. Meanwhile the stronger diagonal alignment hint towards more equally well-trained algorithms.

Interpreting the results of the Principal Component Analysis (\acrshort{PCA}) can be difficult, since the meaning of the components is not necessarily simple to deduct. Since the \acrshort{APS} however is rather uniformly aligned to the diagonal from (0,0) to (1,1), it is not surprising to see the variance by the \acrshort{PCA} generally representing this diagonal. Accordingly, the strong correlation depicted in Figure \ref{fig:PCAdiff} between the first component and the \textit{Difficulty\textsubscript{APS}} metric is to be expected, since it is a representation of this diagonal as shown in Figure \ref{fig:diffGradient}. The different dataset cluster, like the three MovieLens datasets, or the loose cluster of Amazon datasets can subsequently be seen in the \acrshort{PCA} graph as well. Notable outliers in the first components' increasing value, like the Jester dataset, are the best performing datasets. The value of including these datasets in a selection would be great, since they represent the cluster of "solved problems" as shown in Figure \ref{fig:exampleAPS}.
Outliers in the second component could also be a valid candidate for a diverse dataset selection. Since the second component lies orthogonal to the first, seeing some of the datasets with higher \textit{Variance\textsubscript{APS}} is reasonable, since \textit{Variance\textsubscript{APS}} as well lies orthogonal to \textit{Difficulty\textsubscript{APS}} (\acrshort{i.e.} the highly correlative first component). More specifically, the \textit{Variance\textsubscript{APS}} represents the whole orthogonal vector space to \textit{Difficulty\textsubscript{APS}}, while the second component of the \acrshort{PCA} features the one orthogonal dimension to the first component with the highest spread, or variance. Therefore, it is likely that the outliers in the second component would not include all outliers of \textit{Variance\textsubscript{APS}}, making \textit{Variance\textsubscript{APS}} the superior metric to consult when looking for more outliers for more diverse dataset selections. 

When analyzing the \acrshort{APS} through the lens of the \textit{Diversity\textsubscript{APS}}, it becomes apparent how the results enables proper differentiation of datasets. By using sets of two datasets, the \textit{Diversity\textsubscript{APS}} boils down to the euclidean distance between them. This makes a comparison between them quite simple. On the other hand, when using a wider selection of datasets, the second component of \textit{Diversity\textsubscript{APS}} comes into play. While the first component ensures a good coverage of the whole \acrshort{APS}, discouraging clustered sets, the variance component takes the pairwise distances into account. This means for two sets with similar coverage, the variance component encourages an equal distribution of datasets. This effect can be seen, when comparing the sets 2 and 8 of Table \ref{table:resTable}. Set 2, as the strongest set of size 4, features four datasets quite evenly distributed between the "easy" Jester dataset and the "hard" Food dataset, while the \textit{Diversity\textsubscript{APS}} of Set 9 is lower with a score of 0.38, as it contains three clustered datasets, and the Jester dataset as an outlier.
The suggestion of similarity for the MovieLens and Amazon datasets could also be further validated. In sets 7 and 8 of Table \ref{table:resTable}, where they are exclusively featured in dataset selections, a resulting low \textit{Diversity\textsubscript{APS}} could be observed.

A counterintuitive observation can be made, when comparing the \textit{Diversity\textsubscript{APS}} of dataset selections of different sizes. The intuition would suggest that a selection would always become more diverse when adding an additional dataset to the selection, no matter how similar it might be. This is however not reflected in the score of the \textit{Diversity\textsubscript{APS}} metric. Especially when the coverage of the new expanded selection is similar to the old one, the additional dataset has a negative impact on the \textit{Diversity\textsubscript{APS}}. How strong this impact is depends on how the new dataset affects the equal distribution of the selection. With the additional dataset being very close to any of the existing datasets of the selection, the negative impact is strongest, resulting in a lower \textit{Diversity\textsubscript{APS}} for the extended dataset selection. This effect can be seen when comparing the sets 0 and 1 of Table \ref{table:resTable}. Both feature the Jester and Food datasets, categorized by the \acrshort{APS} as the "easiest" and "hardest" datasets to solve with a \textit{Difficulty\textsubscript{APS}} of 0.4 and 0.02 respectively. Set 1 is an extension of set 0 by including the MovieLensLatestSmall dataset, with a \textit{Difficulty\textsubscript{APS}} of 0.2, roughly in the middle between the first two datasets. When put into perspective of the concept of the \acrshort{APS} shown in Figure \ref{fig:exampleAPS} and normalizing the \acrshort{APS} between 0 and 1, the Jester dataset would be considered a "solved problem" of the top right corner, the Food dataset a "true challenge" in the bottom left corner. The additional MovieLensLatestSmall dataset would be located in between (enforced by equal distribution rewarded in \textit{Diversity\textsubscript{APS}}), being representative of the "middle ground" cluster of Figure \ref{fig:exampleAPS}. Adding this new dataset can be considered a diverse and therefore valuable addition to a dataset selection, is however not rewarded properly by the \textit{Difficulty\textsubscript{APS}} metric. This reveals a missing component of the metric, rewarding bigger selections in general, assuming the initially described intention should be reflected in the metric.


\begin{figure}
    \centering
    \includegraphics[width=0.8\linewidth]{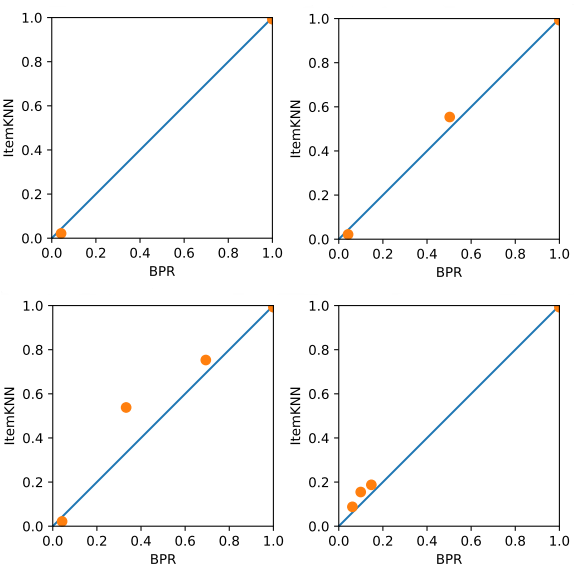}
    \caption{Mini-APS of ItemKNN and BPR featuring dataset selections of Table \ref{table:resTable}. The selections presented are Set 0 in the top left ($\textit{Diversity\textsubscript{APS}}\approx0.47$), Set 1 in top right ($\textit{Diversity\textsubscript{APS}}\approx0.45$), Set 2 in the bottom left ($\textit{Diversity\textsubscript{APS}}\approx0.45$) and Set 8 in bottom right ($\textit{Diversity\textsubscript{APS}}\approx0.38$)}
    \label{fig:selectionHighlights}
\end{figure}

\chapter{Conclusion}
The experiments conducted in the wake of this thesis could further expand the foundation of the Algorithm Performance Space. The work of \cite{beel2024informed} was extended by utilizing hyperparameter optimization, more extensive training and evaluation, and a more in-depth analysis of the \acrshort{APS} by quantifying its results with new metrics. The concept was validated by reproducing observed trends of the previous work, substantiating the hypothesis made of similarity in datasets like the MovieLens ones, and introducing simple metrics to start enabling researchers to justify dataset selections in various ways. Although the limited scope of the experiments diminishes the practical application of the results as a basis for actual dataset selection in research, it revealed further requirements for future work to be done when constructing Algorithm Performance Spaces.
The need for extensive inquiry into every dataset and algorithm featured in the \acrshort{APS} became evident. As the \acrshort{APS} shall be used as a basis for claims based around measured performance, the measurement needs to be as accurate and optimal as possible to be viable. This includes extensive hyperparameter optimization for each algorithm and great time and resource investments, so that bigger datasets or more time-consuming algorithms will not be under-valued. This could include hyperparameter choices being made even on a dataset by dataset basis. Only by respecting the potential of all algorithms used, can the \acrshort{APS} become a truly meaningful tool for research.
This investment only grows in scope, when considering recommendation besides the general problem on implicit feedback data. Assuming a researcher introducing a new algorithm aimed to solve a sequential recommendation problem, arguing dataset selection based upon the \acrshort{APS} as it was constructed here, could become quite meaningless. The need to create \acrshort{APS} for different recommendation problems, featuring different algorithms and datasets, implementing specific pre-processing strategies, becomes quickly apparent.
Keeping the high requirements to create Algorithm Performance Spaces in mind, the effort it takes is worthwhile, as the experiments of this work further validated the theoretical concept. Ideally, the resources only have to be invested once, as the \acrshort{APS} exists as a static entity for the parameters it was created around. It will only be extended by new algorithms as new dimensions of an \acrshort{APS} to be evaluated with the same parameters.

The validity of the \acrshort{APS} is further supported when comparing the Mini-APS of this work and its predecessor from \cite{beel2024informed}. While in the previous work, a strong cluster near 0 performance was observable, in this experiment this cluster, compromised of many of the same datasets, began to dissolve and get higher performances. This can probably be attributed to the \acrshort{HPO} performed and the extended time invested in training. Meanwhile, the same trends of clustered, thematically aligned datasets such as the MovieLens or Amazon datasets could be reproduced. The same is true for certain outliers of "easier" datasets, resulting in greater performance and thus differentiating them from others. In the previous work, some Mini-APS featured trends in the distribution towards certain algorithms, while in this experiment, the distribution of datasets roughly follows the main diagonal. This could indicate that in the previous work, some algorithms did under-perform due to missing \acrshort{HPO} or time investments. The trends seen in the Mini-APS also presented itself in the comparison of the dimensionally reduced graphs using \acrshort{PCA}. Consequently, even though some details of the data pre-processing or training process was adjusted for this experiment, the results still align with expectations made by the introduced concept.

Using the introduced metrics \textit{Difficulty\textsubscript{APS}}, \textit{Variance\textsubscript{APS}} and \textit{Diversity\textsubscript{APS}}, the high-dimensional Algorithm Performance Space becomes tangible, enabling ease of use in justifying strategic dataset selections. While the first two metrics allow for straightforward localization of the dataset inside the \acrshort{APS}, \textit{Diversity\textsubscript{APS}} allows dataset selections to be evaluated based on the differences in their performances. When combining all three metrics, various arguments can be formed to fit and explain a dataset selection based on the requirements set by researchers. Although \textit{Diversity\textsubscript{APS}} is still flawed in some aspect, the metrics already reveal the great potential the \acrshort{APS} has to offer, even in this early iteration.


While with the introduction of Algorithm Performance Spaces and the conducted experiments, the goal of this thesis to provide a new perspective for dataset selection could be reached, more research is required for it to be finalized and to be robust enough to serve researchers as grounds for justifications on dataset selections in the future.

\chapter{Future Work and Limitations}
Since the Algorithm Performance Space is based upon the measured performance of algorithms on datasets, the requirement to extend and improve on these factors seems reasonable. This thesis constructed a rather limited \acrshort{APS}, only featuring five algorithms and 75 datasets. For the high-dimensional space envisioned in the concept, including more algorithms is necessary in future iterations. Additionally, more datasets should be featured as well, since every dataset missing in the \acrshort{APS} is a dataset, that cannot be reasoned for through the lens of the \acrshort{APS}. Furthermore, even more robust training of the algorithms is recommended, the limitations of time and computing resources resulted in a sub-optimal experimental setup. This includes less variable data-splitting like cross-validation splits, more time and resources spent on the training and evaluation, and extended hyperparameter optimization through greater search spaces. By doing so, future iterations of the \acrshort{APS} can include bigger datasets, which largely did not finish or under-perform in this experiment, and more sophisticated algorithms with higher resource requirements or more hyperparameters.

This thesis only featured a singular \acrshort{APS}, based on implicit feedback data and the general recommendation problem, measuring performance in \acrshort{nDCG}@10. Even though this is a popular use-case, researchers focusing on other areas of recommender systems likely require a different \acrshort{APS} for arguing dataset selections. However, the possible number of setups is very high, prioritizing other popular use-cases and creating \acrshort{APS} for them would be of interest. In the case of measuring other performance metrics, this will be quite easy, while creating \acrshort{APS} for other recommendation problems would require more investment, with different sets of datasets and algorithms to be used.

The justification for dataset selections studied in this thesis is based on dataset diversity, through the means of three introduced metrics. These metrics can be altered and extended in the future to allow for more sophisticated arguments to be made. This includes the addition of external information, such as dataset characteristics. Since ultimately the specific use-case of the research should inform dataset selection, only considering dataset diversity might not be the only requirement.

\cleardoublepage
\pagenumbering{Roman}
\setcounter{page}{8}
\phantomsection

\addcontentsline{toc}{chapter}{Bibliography}
\bibliography{bibliography}

\begin{thebibliography}{}

\bibitem[Aiolli, 2013]{aiolli2013efficient}
Aiolli, F. (2013).
\newblock Efficient top-n recommendation for very large scale binary rated datasets.
\newblock In {\em Proceedings of the 7th ACM conference on Recommender systems}, pages 273--280.

\bibitem[Bauer et~al., 2024]{bauer2024exploring}
Bauer, C., Zangerle, E., and Said, A. (2024).
\newblock Exploring the landscape of recommender systems evaluation: Practices and perspectives.
\newblock {\em ACM Transactions on Recommender Systems}, 2(1):1--31.

\bibitem[Beel et~al., 2016]{beel2016towards}
Beel, J., Breitinger, C., Langer, S., Lommatzsch, A., and Gipp, B. (2016).
\newblock Towards reproducibility in recommender-systems research.
\newblock {\em User modeling and user-adapted interaction}, 26:69--101.

\bibitem[Beel and Brunel, 2019]{beel2019data}
Beel, J. and Brunel, V. (2019).
\newblock Data pruning in recommender systems research: Best-practice or malpractice.
\newblock In {\em 13th ACM Conference on Recommender Systems (RecSys)}, volume 2431, pages 26--30.

\bibitem[Beel et~al., 2024]{beel2024informed}
Beel, J., Wegmeth, L., Michiels, L., and Schulz, S. (2024).
\newblock Informed dataset selection with ‘algorithm performance spaces’.
\newblock In {\em Proceedings of the 18th ACM Conference on Recommender Systems}, pages 1085--1090.

\bibitem[Bergstra et~al., 2011]{bergstra2011algorithms}
Bergstra, J., Bardenet, R., Bengio, Y., and K{\'e}gl, B. (2011).
\newblock Algorithms for hyper-parameter optimization.
\newblock {\em Advances in neural information processing systems}, 24.

\bibitem[Chin et~al., 2022]{chin2022datasets}
Chin, J.~Y., Chen, Y., and Cong, G. (2022).
\newblock The datasets dilemma: How much do we really know about recommendation datasets?
\newblock In {\em Proceedings of the Fifteenth ACM International Conference on Web Search and Data Mining}, pages 141--149.

\bibitem[Cremonesi and Jannach, 2021]{cremonesi2021progress}
Cremonesi, P. and Jannach, D. (2021).
\newblock Progress in recommender systems research: Crisis? what crisis?
\newblock {\em AI Magazine}, 42(3):43--54.

\bibitem[Cremonesi et~al., 2010]{cremonesi2010performance}
Cremonesi, P., Koren, Y., and Turrin, R. (2010).
\newblock Performance of recommender algorithms on top-n recommendation tasks.
\newblock In {\em Proceedings of the fourth ACM conference on Recommender systems}, pages 39--46.

\bibitem[Fan et~al., 2024]{fan2024our}
Fan, Y.-c., Ji, Y., Zhang, J., and Sun, A. (2024).
\newblock Our model achieves excellent performance on movielens: what does it mean?
\newblock {\em ACM Transactions on Information Systems}, 42(6):1--25.

\bibitem[Ferrari~Dacrema et~al., 2019]{ferrari2019we}
Ferrari~Dacrema, M., Cremonesi, P., and Jannach, D. (2019).
\newblock Are we really making much progress? a worrying analysis of recent neural recommendation approaches.
\newblock In {\em Proceedings of the 13th ACM conference on recommender systems}, pages 101--109.

\bibitem[Gunawardana et~al., 2012]{gunawardana2012evaluating}
Gunawardana, A., Shani, G., and Yogev, S. (2012).
\newblock Evaluating recommender systems.
\newblock In {\em Recommender systems handbook}, pages 547--601. Springer.

\bibitem[He et~al., 2017]{he2017neural}
He, X., Liao, L., Zhang, H., Nie, L., Hu, X., and Chua, T.-S. (2017).
\newblock Neural collaborative filtering.
\newblock In {\em Proceedings of the 26th international conference on world wide web}, pages 173--182.

\bibitem[Hu et~al., 2008]{hu2008collaborative}
Hu, Y., Koren, Y., and Volinsky, C. (2008).
\newblock Collaborative filtering for implicit feedback datasets.
\newblock In {\em 2008 Eighth IEEE international conference on data mining}, pages 263--272. Ieee.

\bibitem[Koren et~al., 2009]{koren2009matrix}
Koren, Y., Bell, R., and Volinsky, C. (2009).
\newblock Matrix factorization techniques for recommender systems.
\newblock {\em Computer}, 42(8):30--37.

\bibitem[Liang et~al., 2018]{liang2018variational}
Liang, D., Krishnan, R.~G., Hoffman, M.~D., and Jebara, T. (2018).
\newblock Variational autoencoders for collaborative filtering.
\newblock In {\em Proceedings of the 2018 world wide web conference}, pages 689--698.

\bibitem[Pan et~al., 2008]{pan2008one}
Pan, R., Zhou, Y., Cao, B., Liu, N.~N., Lukose, R., Scholz, M., and Yang, Q. (2008).
\newblock One-class collaborative filtering.
\newblock In {\em 2008 Eighth IEEE international conference on data mining}, pages 502--511. IEEE.

\bibitem[Rendle et~al., 2012]{rendle2012bpr}
Rendle, S., Freudenthaler, C., Gantner, Z., and Schmidt-Thieme, L. (2012).
\newblock Bpr: Bayesian personalized ranking from implicit feedback.
\newblock {\em arXiv preprint arXiv:1205.2618}.

\bibitem[Rendle et~al., 2019]{rendle2019difficulty}
Rendle, S., Zhang, L., and Koren, Y. (2019).
\newblock On the difficulty of evaluating baselines: A study on recommender systems.
\newblock {\em arXiv preprint arXiv:1905.01395}.

\bibitem[Sarwar et~al., 2001]{sarwar2001item}
Sarwar, B., Karypis, G., Konstan, J., and Riedl, J. (2001).
\newblock Item-based collaborative filtering recommendation algorithms.
\newblock In {\em Proceedings of the 10th international conference on World Wide Web}, pages 285--295.

\bibitem[Sun et~al., 2022]{sun2022daisyrec}
Sun, Z., Fang, H., Yang, J., Qu, X., Liu, H., Yu, D., Ong, Y.-S., and Zhang, J. (2022).
\newblock Daisyrec 2.0: Benchmarking recommendation for rigorous evaluation.
\newblock {\em IEEE Transactions on Pattern Analysis and Machine Intelligence}, 45(7):8206--8226.

\bibitem[Sun et~al., 2020]{sun2020we}
Sun, Z., Yu, D., Fang, H., Yang, J., Qu, X., Zhang, J., and Geng, C. (2020).
\newblock Are we evaluating rigorously? benchmarking recommendation for reproducible evaluation and fair comparison.
\newblock In {\em Proceedings of the 14th ACM Conference on Recommender Systems}, pages 23--32.

\bibitem[Tyrrell et~al., 2020]{tyrrell2020algorithm}
Tyrrell, B., Bergman, E., Jones, G., and Beel, J. (2020).
\newblock Algorithm-performance personas’ for siamese meta-learning and automated algorithm selection.
\newblock In {\em 7th ICML Workshop on Automated Machine Learning}, volume~1, page~16.

\bibitem[Wu et~al., 2021]{wu2021self}
Wu, J., Wang, X., Feng, F., He, X., Chen, L., Lian, J., and Xie, X. (2021).
\newblock Self-supervised graph learning for recommendation.
\newblock In {\em Proceedings of the 44th international ACM SIGIR conference on research and development in information retrieval}, pages 726--735.

\bibitem[Zhao et~al., 2022]{zhao2022revisiting}
Zhao, W.~X., Lin, Z., Feng, Z., Wang, P., and Wen, J.-R. (2022).
\newblock A revisiting study of appropriate offline evaluation for top-n recommendation algorithms.
\newblock {\em ACM Transactions on Information Systems}, 41(2):1--41.

\bibitem[Zhao et~al., 2021]{zhao2021recbole}
Zhao, W.~X., Mu, S., Hou, Y., Lin, Z., Chen, Y., Pan, X., Li, K., Lu, Y., Wang, H., Tian, C., et~al. (2021).
\newblock Recbole: Towards a unified, comprehensive and efficient framework for recommendation algorithms.
\newblock In {\em proceedings of the 30th acm international conference on information \& knowledge management}, pages 4653--4664.

\end{thebibliography}

\appendix
\chapter{Appendix}
\section{Code Implementation}
The implementation used to conduct the experiment of this thesis, as well as the sources to all datasets, are provided through GitHub at www.github.com/dlay/APS
\section{Experiment Results}
\resizebox{\textwidth}{!}{
\begin{tabular}{lrrrrrrr}
\toprule
 & BPR & ItemKNN & MultiVAE & NeuMF & SGL & Difficulty & Variance \\
Dataset Name &  &  &  &  &  &  &  \\
\midrule
AliEC & 0.0400 & 0.0260 & 0.0414 & 0.0040 & 0.0533 & 0.9671 & 0.0228 \\
Amazon\_Arts\_Crafts\_and\_Sewing & 0.0300 & 0.0426 & 0.0362 & 0.0025 & 0.0361 & 0.9705 & 0.0173 \\
Amazon\_Automotive & NaN & 0.0271 & NaN & NaN & NaN & 0.9729 & NaN \\
Amazon\_CDs\_and\_Vinyl & 0.0541 & 0.0569 & NaN & NaN & NaN & 0.9445 & 0.0028 \\
Amazon\_Cell\_Phones\_and\_Accessories & 0.0114 & 0.0247 & NaN & NaN & NaN & 0.9820 & 0.0133 \\
Amazon\_Clothing\_Shoes\_and\_Jewelry & NaN & 0.0194 & NaN & NaN & NaN & 0.9806 & NaN \\
Amazon\_Digital\_Music & 0.0483 & 0.0752 & 0.0609 & 0.0465 & 0.0573 & 0.9424 & 0.0140 \\
Amazon\_Electronics & NaN & 0.0107 & NaN & NaN & NaN & 0.9893 & NaN \\
Amazon\_Gift\_Cards & 0.0717 & 0.0910 & 0.1003 & 0.0614 & 0.0671 & 0.9217 & 0.0203 \\
Amazon\_Grocery\_and\_Gourmet\_Food & 0.0097 & 0.0263 & NaN & NaN & 0.0040 & 0.9867 & 0.0149 \\
Amazon\_Home\_and\_Kitchen & NaN & 0.0167 & NaN & NaN & NaN & 0.9833 & NaN \\
Amazon\_Industrial\_and\_Scientific & 0.0425 & 0.0609 & 0.0535 & 0.0349 & 0.0581 & 0.9500 & 0.0135 \\
Amazon\_Kindle\_Store & NaN & 0.1345 & NaN & NaN & NaN & 0.8655 & NaN \\
Amazon\_Luxury\_Beauty & 0.0719 & 0.0735 & 0.0677 & 0.0583 & 0.0558 & 0.9346 & 0.0098 \\
Amazon\_Magazine\_Subscriptions & 0.1614 & 0.2612 & 0.2423 & 0.0607 & 0.1854 & 0.8178 & 0.0964 \\
Amazon\_Movies\_and\_TV & 0.0266 & 0.0632 & NaN & NaN & NaN & 0.9551 & 0.0366 \\
Amazon\_Musical\_Instruments & 0.0179 & 0.0268 & 0.0244 & 0.0069 & 0.0254 & 0.9797 & 0.0095 \\
Amazon\_Office\_Products & 0.0260 & 0.0397 & 0.0242 & NaN & 0.0251 & 0.9712 & 0.0079 \\
Amazon\_Patio\_Lawn\_and\_Garden & 0.0098 & 0.0300 & 0.0022 & NaN & 0.0208 & 0.9843 & 0.0157 \\
Amazon\_Pet\_Supplies & 0.0095 & 0.0269 & NaN & NaN & NaN & 0.9818 & 0.0174 \\
Amazon\_Prime\_Pantry & 0.0151 & 0.0204 & 0.0157 & 0.0126 & 0.0251 & 0.9822 & 0.0061 \\
Amazon\_Software & 0.0843 & 0.1313 & 0.1077 & 0.0362 & 0.0941 & 0.9093 & 0.0427 \\
Amazon\_Sports\_and\_Outdoors & 0.0068 & 0.0263 & NaN & NaN & NaN & 0.9834 & 0.0195 \\
Amazon\_Tools\_and\_Home\_Improvement & 0.0017 & 0.0209 & NaN & NaN & NaN & 0.9887 & 0.0192 \\
Amazon\_Toys\_and\_Games & 0.0203 & 0.0492 & NaN & NaN & NaN & 0.9652 & 0.0289 \\
Amazon\_Video\_Games & 0.0366 & 0.0340 & 0.0334 & 0.0007 & 0.0379 & 0.9715 & 0.0155 \\
Anime & 0.3819 & 0.3744 & 0.3880 & NaN & NaN & 0.6186 & 0.0091 \\
BeerAdvocate & 0.1135 & 0.1270 & 0.1226 & 0.0990 & 0.1155 & 0.8845 & 0.0130 \\
Behance & 0.0448 & 0.0944 & 0.0618 & NaN & 0.0541 & 0.9362 & 0.0261 \\
BookCrossing & 0.0553 & 0.0776 & 0.0540 & 0.0403 & 0.0653 & 0.9415 & 0.0172 \\
CiaoDVD & 0.0568 & 0.0703 & 0.0769 & 0.0657 & 0.0808 & 0.9299 & 0.0118 \\
CiteULike-a & 0.1461 & 0.2084 & 0.1909 & 0.1377 & 0.1606 & 0.8313 & 0.0372 \\
CosmeticsShop & 0.0758 & 0.1165 & NaN & 0.0239 & NaN & 0.9279 & 0.0617 \\
\bottomrule
\end{tabular}
}

\resizebox{\textwidth}{!}{
\begin{tabular}{lrrrrrrr}
\toprule
 & BPR & ItemKNN & MultiVAE & NeuMF & SGL & Difficulty & Variance \\
Dataset Name &  &  &  &  &  &  &  \\
\midrule
DeliveryHeroSE & 0.1489 & 0.1652 & 0.1324 & 0.1311 & 0.1602 & 0.8524 & 0.0192 \\
DeliveryHeroSG & NaN & 0.1681 & NaN & NaN & NaN & 0.8319 & NaN \\
DeliveryHeroTW & 0.0589 & 0.1835 & NaN & NaN & NaN & 0.8788 & 0.1246 \\
DoubanBook & 0.1183 & 0.2006 & 0.1593 & 0.0545 & 0.1664 & 0.8602 & 0.0681 \\
DoubanMovie & 0.1733 & 0.2051 & NaN & 0.1244 & NaN & 0.8324 & 0.0538 \\
DoubanMusic & 0.1188 & 0.1942 & 0.1343 & 0.0876 & 0.0333 & 0.8864 & 0.0737 \\
Epinions & 0.0722 & 0.3757 & NaN & NaN & NaN & 0.7760 & 0.3035 \\
FilmTrust & 0.3367 & 0.3656 & 0.3771 & 0.3662 & 0.2474 & 0.6614 & 0.0578 \\
Food & 0.0205 & 0.0105 & 0.0236 & 0.0005 & 0.0132 & 0.9863 & 0.0112 \\
FourSquareNYC & 0.0721 & 0.0763 & 0.0736 & 0.0736 & 0.0594 & 0.9290 & 0.0071 \\
FourSquareTokyo & 0.1521 & 0.1729 & 0.1581 & 0.1562 & 0.1716 & 0.8378 & 0.0114 \\
Globo & 0.1399 & 0.1677 & NaN & NaN & NaN & 0.8462 & 0.0278 \\
GoodReadsComics & 0.2463 & 0.4046 & NaN & NaN & NaN & 0.6746 & 0.1583 \\
GoogleLocalAlaska & 0.0967 & 0.1074 & 0.1103 & 0.0206 & 0.0800 & 0.9170 & 0.0414 \\
GoogleLocalDelaware & 0.0877 & 0.0997 & 0.0915 & NaN & NaN & 0.9070 & 0.0080 \\
GoogleLocalDistrictOfColumbia & 0.0733 & 0.0885 & 0.0811 & 0.0301 & NaN & 0.9318 & 0.0305 \\
GoogleLocalMontana & 0.0785 & 0.0872 & 0.0884 & NaN & 0.0129 & 0.9332 & 0.0392 \\
GoogleLocalVermont & 0.1005 & 0.1038 & 0.1108 & 0.0262 & 0.1178 & 0.9082 & 0.0387 \\
Gowalla & NaN & 0.0976 & NaN & NaN & NaN & 0.9024 & NaN \\
Jester & 0.4854 & 0.4818 & 0.5023 & 0.4907 & 0.4584 & 0.5163 & 0.0193 \\
LastFM & 0.2423 & 0.2686 & 0.2576 & 0.2518 & 0.2501 & 0.7459 & 0.0120 \\
LearningFromSets & 0.3932 & 0.4306 & 0.3960 & 0.3757 & 0.4047 & 0.6000 & 0.0243 \\
LibraryThing & 0.0462 & 0.0968 & 0.0262 & 0.0183 & 0.0547 & 0.9516 & 0.0371 \\
MIND-Small & 0.0430 & 0.0489 & 0.0485 & NaN & NaN & 0.9532 & 0.0039 \\
MarketBiasModcloth & 0.0967 & 0.0862 & 0.1026 & 0.1051 & 0.0594 & 0.9100 & 0.0216 \\
ModCloth & 0.1330 & 0.1849 & 0.1582 & 0.1178 & 0.1237 & 0.8565 & 0.0337 \\
MovieLens100k & 0.2680 & 0.2812 & 0.2804 & 0.2716 & 0.2379 & 0.7322 & 0.0198 \\
MovieLens1m & 0.2871 & 0.2868 & 0.2965 & 0.2566 & 0.2960 & 0.7154 & 0.0178 \\
MovieLensLatestSmall & 0.2441 & 0.2688 & 0.2610 & 0.2323 & 0.2016 & 0.7584 & 0.0326 \\
MovieTweetings & 0.0977 & 0.1330 & 0.1277 & 0.0914 & 0.1217 & 0.8857 & 0.0226 \\
Netflix & NaN & 0.2579 & NaN & NaN & NaN & 0.7421 & NaN \\
RateBeer & 0.1598 & 0.1891 & 0.1834 & 0.1276 & 0.0199 & 0.8640 & 0.0788 \\
Rekko & 0.0926 & 0.1076 & 0.0987 & 0.0895 & 0.1026 & 0.9018 & 0.0092 \\
RentTheRunway & 0.0109 & 0.0063 & 0.0203 & 0.0183 & 0.0249 & 0.9839 & 0.0093 \\
Retailrocket & 0.1413 & 0.1659 & 0.1554 & 0.1120 & 0.1241 & 0.8603 & 0.0278 \\
TaFeng & 0.0487 & 0.0969 & 0.0617 & 0.0198 & 0.0652 & 0.9415 & 0.0341 \\
Twitch100k & 0.2118 & 0.2786 & 0.2853 & NaN & NaN & 0.7414 & 0.0490 \\
Yelp & 0.0248 & 0.0320 & NaN & NaN & NaN & 0.9716 & 0.0072 \\
\bottomrule
\end{tabular}
}

\section{Datasets Metadata}

\resizebox{\linewidth}{!}{
\begin{tabular}{llrrrl}
\toprule
 & Dataset Name & Interactions & Users & Items \\
\midrule
0 & AliEC & 252092 & 30461 & 16475 \\
1 & Amazon\_Arts\_Crafts\_and\_Sewing & 299121 & 17200 & 34849 \\
2 & Amazon\_Automotive & 1130686 & 58718 & 131694 \\
3 & Amazon\_Books & 21301776 & 597748 & 1514288 \\
4 & Amazon\_CDs\_and\_Vinyl & 1075615 & 59934 & 87712 \\
5 & Amazon\_Cell\_Phones\_and\_Accessories & 580559 & 29537 & 82276 \\
6 & Amazon\_Clothing\_Shoes\_and\_Jewelry & 7130297 & 274068 & 811738 \\
7 & Amazon\_Digital\_Music & 104431 & 8582 & 10623 \\
8 & Amazon\_Electronics & 4406114 & 121928 & 490998 \\
9 & Amazon\_Fashion & 2482 & 8 & 314 \\
10 & Amazon\_Gift\_Cards & 2792 & 144 & 430 \\
11 & Amazon\_Grocery\_and\_Gourmet\_Food & 722704 & 30609 & 83740 \\
12 & Amazon\_Home\_and\_Kitchen & 4538688 & 147204 & 528263 \\
13 & Amazon\_Industrial\_and\_Scientific & 34809 & 2871 & 5061 \\
14 & Amazon\_Kindle\_Store & 1767927 & 84671 & 111033 \\
15 & Amazon\_Luxury\_Beauty & 13159 & 731 & 1349 \\
16 & Amazon\_Magazine\_Subscriptions & 1000 & 74 & 154 \\
17 & Amazon\_Movies\_and\_TV & 2281782 & 48208 & 209649 \\
18 & Amazon\_Musical\_Instruments & 147173 & 7414 & 18177 \\
19 & Amazon\_Office\_Products & 487519 & 20193 & 62738 \\
20 & Amazon\_Patio\_Lawn\_and\_Garden & 425747 & 21897 & 56991 \\
21 & Amazon\_Pet\_Supplies & 1234754 & 32088 & 147033 \\
22 & Amazon\_Prime\_Pantry & 101636 & 4311 & 10554 \\
23 & Amazon\_Software & 3298 & 342 & 482 \\
24 & Amazon\_Sports\_and\_Outdoors & 1799524 & 77781 & 216349 \\
25 & Amazon\_Tools\_and\_Home\_Improvement & 1333795 & 55958 & 159584 \\
26 & Amazon\_Toys\_and\_Games & 1237687 & 61125 & 143058 \\
27 & Amazon\_Video\_Games & 291985 & 12455 & 33625 \\
28 & Anime & 6686956 & 66464 & 8003 \\
29 & BeerAdvocate & 1189084 & 13391 & 18529 \\
30 & Behance & 687070 & 23724 & 29794 \\
31 & BookCrossing & 77527 & 4923 & 6458 \\
32 & CiaoDVD & 17263 & 1315 & 1372 \\
33 & CiteULike-a & 200180 & 5536 & 15429 \\
34 & CosmeticsShop & 1081605 & 64671 & 25159 \\
\bottomrule
\end{tabular}
}

\resizebox{\linewidth}{!}{
\begin{tabular}{llrrrl}
\toprule
 & Dataset Name & Interactions & Users & Items \\
\midrule
35 & DeliveryHeroSE & 347987 & 36088 & 17932 \\
36 & DeliveryHeroSG & 2329565 & 196188 & 120789 \\
37 & DeliveryHeroTW & 2632313 & 256630 & 163308 \\
38 & DoubanBook & 809697 & 23585 & 23386 \\
39 & DoubanMovie & 6345859 & 58021 & 29621 \\
40 & DoubanMusic & 853574 & 17834 & 25067 \\
41 & DoubanShort & 411835 & 54409 & 28 \\
42 & Epinions & 11935999 & 42708 & 478786 \\
43 & FilmTrust & 13201 & 926 & 178 \\
44 & Food & 498607 & 16043 & 37851 \\
45 & FourSquareNYC & 40825 & 1062 & 3895 \\
46 & FourSquareTokyo & 128530 & 2287 & 7055 \\
47 & Globo & 2482163 & 157926 & 11832 \\
48 & GoodReadsComics & 3005157 & 88907 & 46869 \\
49 & GoogleLocalAlaska & 536962 & 36192 & 8141 \\
50 & GoogleLocalDelaware & 896048 & 59338 & 10325 \\
51 & GoogleLocalDistrictOfColumbia & 687318 & 60289 & 7090 \\
52 & GoogleLocalMontana & 967230 & 62021 & 14456 \\
53 & GoogleLocalVermont & 368772 & 28220 & 7090 \\
54 & Gowalla & 2018421 & 64115 & 164532 \\
55 & Jester & 42813 & 2554 & 136 \\
56 & LastFM & 71355 & 1859 & 2823 \\
57 & LearningFromSets & 262642 & 854 & 6517 \\
58 & LibraryThing & 531962 & 18510 & 32991 \\
59 & MarketBiasModcloth & 28508 & 1880 & 652 \\
60 & MillionSong & 48146077 & 1019291 & 285048 \\
61 & MIND-Small & 2161132 & 85570 & 19602 \\
62 & ModCloth & 7172 & 1071 & 236 \\
63 & MovieLens1m & 574376 & 6034 & 3125 \\
64 & MovieLens100k & 54413 & 938 & 1008 \\
65 & MovieLensLatestSmall & 53371 & 602 & 2412 \\
66 & MovieTweetings & 563309 & 20643 & 8810 \\
67 & Netflix & 56879880 & 463435 & 17721 \\
68 & RateBeer & 1765305 & 10644 & 36129 \\
69 & Rekko & 219610 & 19540 & 3796 \\
70 & RentTheRunway & 33648 & 3985 & 2817 \\
71 & Retailrocket & 248891 & 22890 & 18269 \\
72 & TaFeng & 709356 & 26039 & 15483 \\
73 & Twitch100k & 1228857 & 76411 & 27246 \\
74 & Yelp & 2428509 & 174840 & 77319 \\
\bottomrule
\end{tabular}
}

\clearpage




\end{document}